\documentclass[aps,prl,reprint,amsmath,amssymb,floatfix,longbibliography,superscriptaddress]{revtex4-1}

\usepackage{graphicx}
\usepackage{amsmath,amssymb,bbold,bm,color}
\usepackage{float}
\usepackage{epstopdf}
\usepackage{hyperref}
\usepackage[dvipsnames]{xcolor}
\usepackage[normalem]{ulem} 

\newcommand{\cT}{{\cal T}}

\newcommand{\bee}{\begin{equation}}
\newcommand{\ee}{\end{equation}}
\newcommand{\bra}[1]{\langle #1 |}
\newcommand{\ket}[1]{| #1 \rangle}

\hypersetup{colorlinks=true,linkcolor=blue,citecolor=blue,urlcolor=blue}

\begin{document}
\title{Revival dynamics in a traversable wormhole}

\author{Stephan Plugge}
\email{plugge@phas.ubc.ca}
\affiliation{Department of Physics and Astronomy \& Stewart Blusson Quantum Matter Institute, University of British Columbia, Vancouver BC, Canada V6T 1Z4}
\author{\'{E}tienne Lantagne-Hurtubise}
\email{lantagne@phas.ubc.ca}
\affiliation{Department of Physics and Astronomy \& Stewart Blusson Quantum Matter Institute, University of British Columbia, Vancouver BC, Canada V6T 1Z4}
\affiliation{Kavli Institute for Theoretical Physics, University of California Santa Barbara, CA 93106, USA}
\author{Marcel Franz}
\affiliation{Department of Physics and Astronomy \& Stewart Blusson Quantum Matter Institute, University of British Columbia, Vancouver BC, Canada V6T 1Z4}

\date{\today}

\begin{abstract} 
Quantum effects can stabilize wormhole solutions in general relativity, allowing information and matter to be transported between two connected spacetimes. Here we study the revival dynamics of signals sent between two weakly coupled quantum chaotic systems, represented as identical Sachdev-Ye-Kitaev models, that realize holographically a traversable wormhole in anti-de Sitter spacetime AdS$_2$ for large number $N$ of particles. In this limit we find clear signatures of wormhole behavior:
an excitation created in one system is quickly scrambled under its unitary dynamics, and is reassembled in the other system after a characteristic time consistent with holography predictions. This leads to revival oscillations that at low but finite temperature decay as a power-law in time. For small $N$ we also observe revivals and show that they arise from a different, non-gravitational mechanism. 
\end{abstract}

\date{\today}
\maketitle

General relativity allows for wormhole solutions connecting
two approximately flat space-times -- the Einstein-Rosen bridge
\cite{Einstein35}. Unfortunately,
such wormholes are known to be unstable and cannot be used to transmit matter or
information. The ER bridge solution can be stabilized
by the insertion of exotic matter with negative rest energy, but such matter is not available classically \cite{Thorne88}. In a
remarkable development, recent work in quantum gravity demonstrated that quantum effects
can substitute for the exotic matter and stabilize a \emph{traversable wormhole} \cite{Gao2017,Yang2017,Maldacena2018,Bak2018,GaoLiu2019,Marolf2019,Bak2019}.
From a practical standpoint it is unlikely that genuine quantum
gravity effects, such as the traversable wormhole formation, can be
probed experimentally in the foreseeable future.  One can imagine,
however, testing these effects in quantum systems that are related to gravity by holographic dualities -- thus enabling, at least in principle, explorations of otherwise inaccessible physical  phenomena in tabletop experiments. 

An example of this duality, notable for its simplicity and wide
appeal, is the Sachdev-Ye-Kitaev (SYK)
model \cite{SY1996,Kitaev2015},
which at low energies is believed to be holographically dual to a near-extremal black hole with an AdS$_2$ horizon
\cite{Kitaev2015,Sachdev2015,Maldacena2016}. The model shows extensive residual entropy, emergent (weakly
broken) conformal invariance and is maximally chaotic \cite{Maldacena2016b}, all properties shared with black holes. At the
same time the SYK model is sufficiently simple to be analytically tractable in the limit of a large number $N$ of particles. There now exist several proposed physical realizations of the SYK model and its variants in atomic and solid-state systems \cite{Danshita2017,Pikulin2017,Alicea2017,achen2018,Rozali2018,Altland2019},
making it a potential platform for laboratory studies of quantum dynamics of black holes.

In this Letter we study a pair of weakly coupled SYK models  which, according to recent work by Maldacena and Qi \cite{Qi2018},
furnishes 
a holographic realization of an eternal traversable wormhole with an AdS$_2$ throat.  The model exhibits many interesting properties~\cite{Garcia2019,Maldacena2019,Chen2019,Alet2020} and its
physical realization based on the proposed SYK platforms has been discussed~\cite{Etienne2019}.  
It is defined by Hamiltonian
\begin{equation} \label{m1}
H = H_L^{\rm SYK} + H_R^{\rm SYK}  + i \mu \sum_j \chi_L^j \chi_R^j
\end{equation}
where  $H_\alpha^{\rm SYK} $ describe the SYK models 
\begin{equation}\label{m2}
H_\alpha^{\rm SYK}  = \sum_{i<j<k<l} J_{ijkl} \chi_{\alpha}^i \chi_{\alpha}^j \chi_{\alpha}^k \chi_{\alpha}^l~,
\end{equation}
and $\alpha=L,R$ refer to the ``left" and ``right" side of the wormhole. Each contains $N$ Majorana operators respecting $(\chi_{\alpha}^j)^\dagger = \chi_{\alpha}^j$ and $\{ \chi_{\alpha}^i , \chi_{\beta}^j \} = \delta^{ij} \delta_{\alpha \beta}$. The real-valued random couplings $J_{ijkl}$ are drawn from a Gaussian distribution with $ \overline{J_{ijkl}} = 0,~\overline{J_{ijkl}^2} = \frac{3! J^2}{N^3} $
and are, crucially, identical for $\alpha=L,R$.

The Maldacena-Qi (MQ) model Eq.~\eqref{m1} is believed to be dual to a traversable wormhole for weak coupling $\mu \ll J$ and low temperatures $T \lesssim \mu$~\cite{Qi2018}. The key property that makes it a wormhole is the following. Starting in the ground state $|\Psi_0\rangle$ of the model,
imagine creating an excitation on the right side $\chi^j_R |\Psi_0\rangle$. In the absence of coupling ($\mu = 0$), the excitation will rapidly dissipate due to the chaotic nature of the SYK model~\cite{Kitaev2015,Maldacena2016}. Quantum information contained in the excitation will be {\em scrambled} among a large number of states in the Hilbert space of $H_R^{\rm SYK}$, becoming effectively lost to all simple observables.
Remarkably, when the two systems are coupled by a small $\mu$, the scrambled excitation is instead transported to the left side where it is `unscrambled' to its original form: after a characteristic time $t_\mathrm{re}$, the state of the system becomes close to $\chi^j_L |\Psi_0\rangle$. In quantum gravity language, the particle has passed through the wormhole.

\begin{figure*}
	\includegraphics[width=0.95\textwidth]{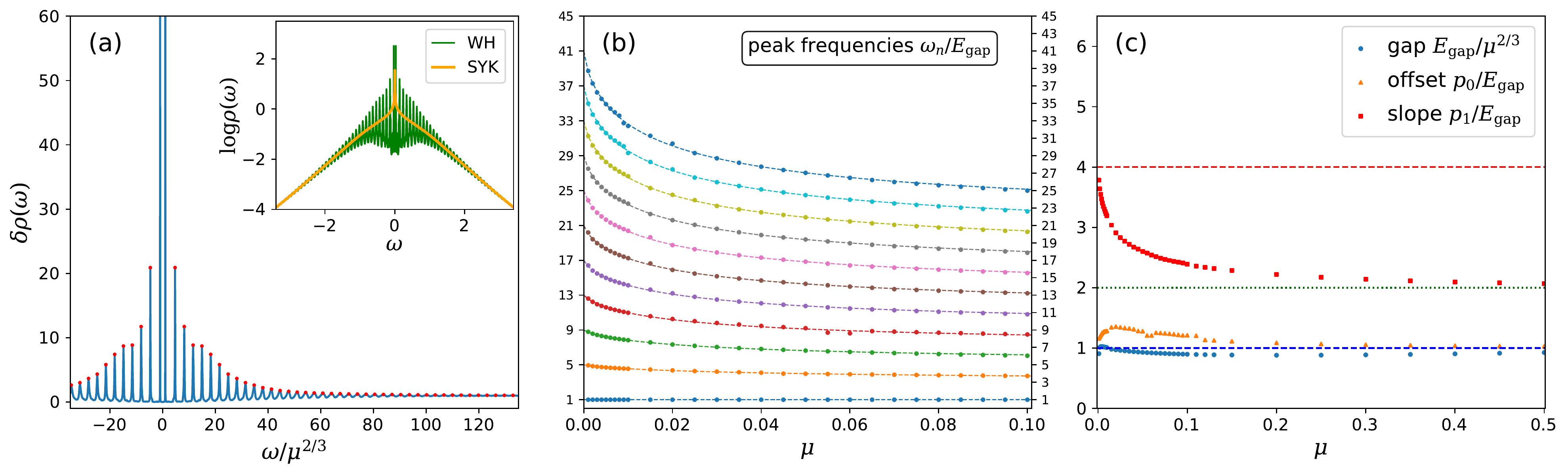}
	\caption{
		Spectral features of the MQ model obtained from numerical solution of the SD equations~\eqref{m5} in real time and at temperatures $T \ll \mu$ below the Hawking-Page transition~\cite{Qi2018} ($T=2\cdot 10^{-4}$ for $\mu\geq 0.01$, $T=5\cdot 10^{-5}$ for $\mu< 0.01$).
		(a) inset: spectral function for SYK  ($\rho_0,~ \mu=0$) and wormhole ($\rho_{LL},~\mu=0.004$).
		Main panel: relative spectral weight $\delta \rho(\omega) = \rho_{LL}/\rho_0$. Red dots indicate positions of the dominant peaks analyzed in (b,c).
		(b) peak frequencies $\omega_n >0$, extracted from peak positions of $\delta \rho(\omega)$ in (a). For $\mu\to 0$ and $\omega_n \ll J$ we see a clear approach to the conformal tower $\omega_n = E_{\rm gap}(4n+1)$~\cite{Qi2018}. The dashed lines are a simple fit $Y_n(\mu)$, see text.
		(c) spectral gap $E_{\rm gap} = \omega_0$, and offset $p_0$ and slope $p_1$ of linear fits $\omega_n \sim p_0 + p_1 n$.
		\label{fig:SDspec}}
\end{figure*}

While this revival property has been predicted based on the gravity interpretation of the MQ model~\cite{Qi2018}, a direct observation within the quantum mechanical description of Eqs.~\eqref{m1}-\eqref{m2} is yet to be achieved. This is the main goal of the present work. To this end we employ two complementary approaches.
First, numerical exact diagonalization (ED) of the Hamiltonian \eqref{m1} yields the complete set of many-body
eigenstates $|\Psi_n\rangle$ with energy $E_n$. Any observable or correlation function can then be computed exactly; however, ED is limited to small systems with $2N \leq 32$.
The second approach consists of solving the saddle-point equations for the averaged fermion propagator
\begin{equation}\label{m4}
G_{\alpha\beta}(\tau_1,\tau_2)={1\over N}
\sum_j\langle\cT \chi_\alpha^j(\tau_1)\chi_\beta^j(\tau_2)\rangle,
\end{equation}
where $\cT$ denotes imaginary-time ordering and $\langle \cdots
\rangle$ stands for thermal average. For the SYK model this approach is asymptotically exact in the limit of large $N$.

\emph{Large-N solution.}-- Under the assumption of time-translation invariance,
$G_{\alpha\beta}(\tau_1,\tau_2) =G_{\alpha\beta}(\tau_1-\tau_2)$, and
the mirror symmetry between $L$ and $R$ systems which constrains
$G_{RR}(\tau)=G_{LL}(\tau)$ and $G_{RL}(\tau)=-G_{LR}(\tau)$, the saddle-point equations of the MQ model can be reduced to a single pair of self-consistent equations~\cite{SuppMat},
\begin{eqnarray}\label{m5}
G_+(i\omega_n)&=&\left[i\omega_n-\mu-\Sigma_+(i\omega_n)\right]^{-1},\\
\Sigma_+(\tau)&=&-{1\over 4}J^2\left[3G_+^2 (\tau)G_+(-\tau)+G_+^3
  (-\tau)\right].
\nonumber
\end{eqnarray}
Here we defined $G_\pm(\tau)=G_{LL}(\tau)\pm iG_{LR}(\tau)$ and similar for the self energies $\Sigma_\pm(\tau)$ (henceforth, we set $J=1$). The advantage of this
representation is that the equations for $G_+$ and $G_-$ decouple. Since the propagators are related by $G_\pm(\tau)=-G_\mp(-\tau)$, solving Eqs.~\eqref{m5} yields both $G_\pm(\tau)$ which can be used to reconstruct
the full $G_{\alpha\beta}(\tau)$.

To access dynamical properties of the MQ model, we switch to real-time representation of the SD equations in Eq.~\eqref{m5} and employ a weighted-iteration scheme~\cite{Qi2018,Etienne2019} to find self-consistent solutions~$(G, \Sigma)$~\cite{SuppMat}. In the limit of weak coupling $\mu$ where wormhole effects are presumed to arise, it is necessary to employ both a large frequency cutoff $\omega_{\mathrm{max}} \gg J$ and small spacing $\delta\omega \ll \mu$ to resolve fine details of the spectral function \emph{and} capture slowly decaying tails at high energies that are characteristic of SYK physics. This makes numerics challenging
\footnote{In order to observe revivals, it is crucial to extract more than just an energy gap from saddle-point equations or exact diagonalization. A single-scale expression $G_{LL}(\tau) \sim e^{-E_\mathrm{gap}\tau} \to G_{LL}(t) \sim e^{-iE_\mathrm{gap}t}$~\cite{Qi2018,Garcia2019} does \emph{not} predict revivals, cf. Eqs.~\eqref{m7}. In imaginary-time formulation, revivals associated with higher energy scales (conformal tower) are hidden in the early-time, non-exponential transient behavior of $G_{LL}(\tau)$.}.
By implementing the symmetries in Eq.~\eqref{m5} we are able to go beyond earlier works~\cite{Qi2018,Garcia2019}.

\emph{Spectral function.}-- In Fig.~\ref{fig:SDspec} we analyze the spectral function $\rho_{LL}(\omega) = -\frac{1}{\pi} {\rm Im} G_{LL}^{\rm ret}(\omega)$ of the MQ model obtained from the large-$N$ saddle-point solution. A key observation in Fig.~\ref{fig:SDspec}a is the series of evenly-spaced spectral peaks at frequencies $\omega_n$ shown in Fig.~\ref{fig:SDspec}b.
For small couplings $\mu$, holography predicts the emergence of two conformal ``towers" of states at low energies~\cite{Qi2018}
\begin{equation}\label{m6}
E_n^{\rm conf} = \epsilon \left(\Delta + n \right),~~ E_n^{\rm bg} = \epsilon \sqrt{2(1-\Delta)} \left(n + \frac12\right),
\end{equation}
where $\Delta =1/4$ is the fermion scaling dimension. These towers correspond to conformal excitations and a ``boundary graviton'' in the gravity interpretation of the MQ model~\cite{Qi2018}, respectively. From the SD equations~\eqref{m5} and scaling arguments one expects $\epsilon \sim \mu^{2/3}$, leading to a spectral gap $E_{\rm gap} \sim \Delta \mu^{2/3}$ and $E_n^{\rm conf} \simeq E_{\rm gap}(4n+1)$. At large $\mu$ the system instead has a trivial (non-interacting) harmonic oscillator spectrum, $E^{\rm triv}_n \simeq E_{\rm gap}(2n+1)$.
Our data in Fig.~\ref{fig:SDspec}b-c confirms this for the quantum-mechanical side of the MQ model.
Fig.~\ref{fig:SDspec}b also shows fits to an empirical function $Y_n(\mu) = E_{\rm gap} \left(1+2n\frac{2+c_n\mu^{\nu_n}}{1+c_n\mu^{\nu_n}}\right)$ with free parameters $c_n, \nu_n$, that captures the conformal and trivial limits and matches the data well.
For Fig.~\ref{fig:SDspec}c, at each $\mu$ we fit the lowest eight peaks of the series $\omega_n$ of Fig.~\ref{fig:SDspec}b by a simple linear form $\omega_n \sim p_0 +p_1 n$. As an indicator of the inverse scaling dimension $\Delta^{-1}$, the ratio $p_1/E_{\rm gap}$ changes from $2 \to 4$ as $\mu$ gets smaller. The spectral gap is close to the value $E_{\rm gap} = \mu^{2/3}$~\cite{Qi2018}. Note that the offset $p_0$ does not match the gap $\omega_0$ since the peak series $\omega_n$ is skewed and deviates from a purely linear behavior at high frequencies.
Further, we find no evidence for the ``boundary graviton'' states among the spectral peaks in $\rho_{LL}(\omega)$
\footnote{The boundary graviton states are expected to occur at higher order in the $1/N$ expansion, and thus do not appear in our saddle-point calculations. We thank J. Maldacena and A. Milekhin for pointing this out to us.
}.

\emph{Revival dynamics.} -- Revival dynamics are captured by the transmission amplitudes between the two systems
\begin{equation}\label{m7}
T_{\alpha\beta}(t) = 2|G_{\alpha\beta}^>(t)| \: , \; G^>_{\alpha \beta}(t) = \frac{\theta(t)}{N} \sum_j \langle \chi^j_\alpha(t) \chi^j_\beta(0) \rangle .
\end{equation}
The transmission $T_{\alpha\beta}^2(t)$ then reflects the probability to recover $\chi_\alpha^j(t)$ after inserting $\chi_\beta^j(0)$ initially, averaged over all modes $j$. 

\begin{figure}[t]	\includegraphics[width=0.95\columnwidth]{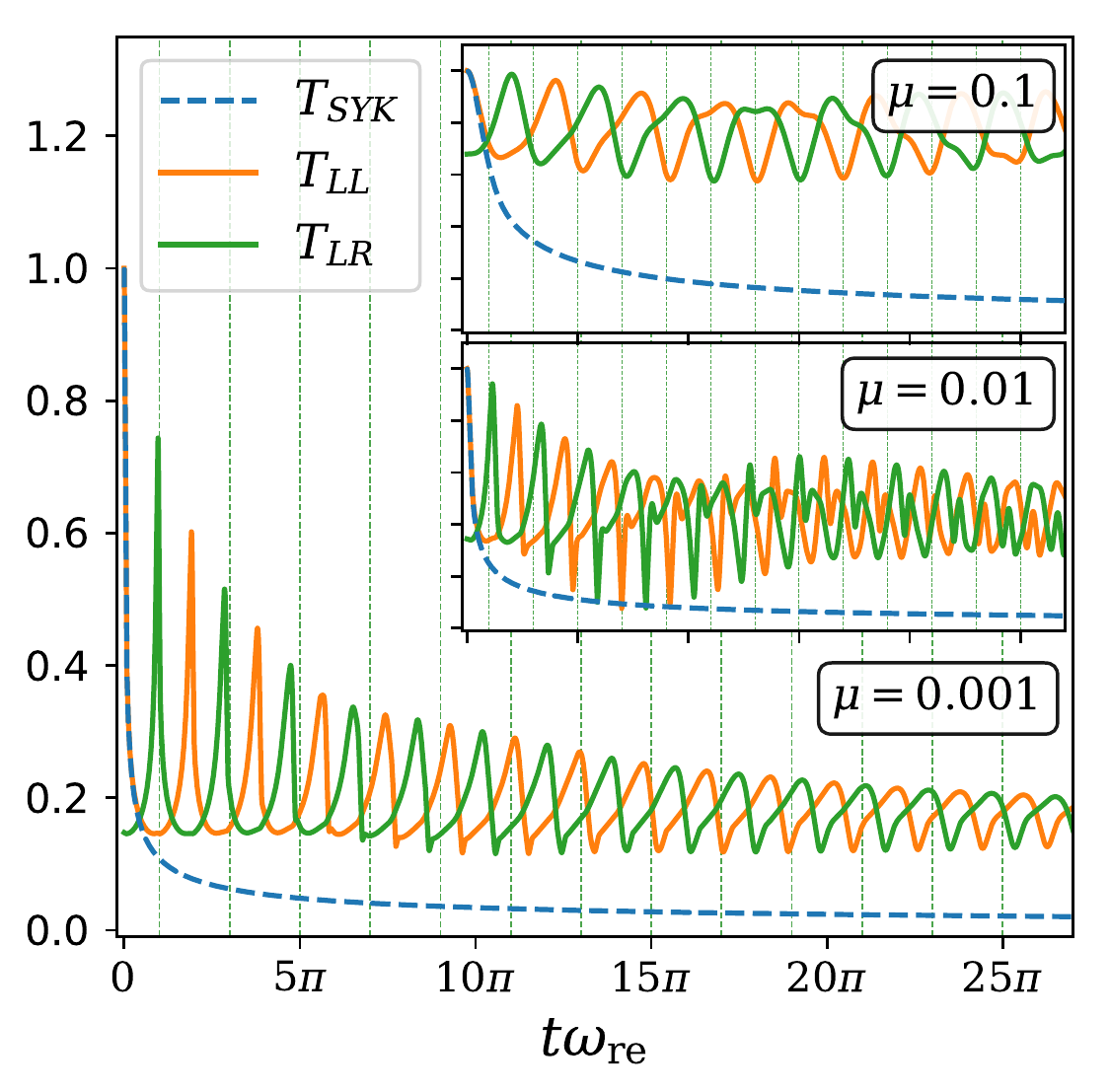}
	\caption{
		Revival dynamics in transmission amplitudes $T_{LL,LR}$ [Eq.~\eqref{m7}] at temperature $T < \mu$; $T_{\rm SYK} = T_{LL}(\mu=0)$ refers to an uncoupled SYK model.
		We show data for various $\mu$, with time axes rescaled by $\omega_{\rm re} = p_1/2\pi$, cf. Fig.~\ref{fig:SDspec}. Vertical  lines indicate times $t_{\mathrm{re},n} = (2n+1)\pi /\omega_{\rm re}$ for which an excitation $\chi^j_R \ket{\Psi_0}$ is expected to re-assemble on the left side.
	\label{fig:SDrev}}
\end{figure}

In Fig.~\ref{fig:SDrev} we show $T_{\alpha \beta}(t)$ obtained from the numerical solution of the SD equations. For $\mu=0$, we recover the usual SYK power-law $T_\mathrm{SYK}(t) = 2|G^>_\mathrm{SYK}(t)| \sim t^{-1/2}$. 
For small $\mu > 0$, $T_{LL}(t)$ initially follows closely the SYK result. Then, at time $t_\mathrm{re}$, a sharp peak in $T_{LR}(t)$ indicates that the excitation has traversed to the other side.
The situation at $t=t_\mathrm{re}$ is similar to the initial configuration with $L$ and $R$ reversed -- thus at $t = 2t_\mathrm{re}$ we expect a recurrence of the excitation at its original position, indicated by a peak at $T_{LL}(2t_\mathrm{re})$, and so on.
As found in Fig.~\ref{fig:SDrev}, the characteristic frequency $\omega_\mathrm{re} = \frac{p_1}{2\pi} \sim\mu^{2/3}$ for such oscillations is consistent with the prediction from holography~\cite{Qi2018,SuppMat}. It is much larger than the naive guess $\omega_\mathrm{re} \sim\mu$, indicating that the chaotic SYK interactions assist the transmission of information between the two sides of the wormhole. 
For small $\mu$ the oscillations decay as a power-law in time with an exponent approaching $\frac{1}{2}$~\cite{SuppMat}, indicative of an SYK-like envelope to the revivals.
This shows that at finite temperature the MQ model contains corrections that allow particles to thermalize; the effect is yet to be discussed in the gravitational context.
At high temperatures, the system transitions from a wormhole to two decoupled black holes~\cite{Qi2018} where revivals are absent.
Further, the observed revivals can be contrasted with the large $\mu$ limit, which is not dual to a wormhole. A simple calculation gives $T_{\alpha\beta}(t)\equiv 1$ in the ground state of the system for $\mu \gg 1$, which is reflected in the suppressed oscillation amplitudes for large $\mu$, cf. Fig.~\ref{fig:SDrev}.

Note that the Fourier transform of the transmission, $F_{\alpha \beta}(\omega) = \int dt e^{i \omega t} T_{\alpha \beta}^2 (t)$, can be written as a spectral auto-correlation function~\cite{SuppMat}
\begin{equation}
    F_{\alpha \beta}(\omega) =
    8\pi\int_0^\infty d \omega' \rho_{\alpha \beta} (\omega') \rho_{\alpha \beta}(\omega' + |\omega|).
    \label{eq:conv}
\end{equation}
Sharp revivals thus rely on a series of evenly-spaced peaks in $F_{\alpha\beta}(\omega)$, originating from strong peak-spacing correlations in the spectral functions $\rho_{\alpha \beta}(\omega)$, cf. Fig.~\ref{fig:SDspec}. The ``beating'' in Fig.~\ref{fig:SDrev} is due to deviations from linearity in the peak series $\omega_n$.
The transmission does not decay to zero between peaks -- this finite value persists up to times exponentially long in the inverse temperature~$\beta$, which sets the width of the $\omega=0$ peak in $F_{\alpha \beta}(\omega)$~\cite{QiZhang2020,SuppMat}. Then the system eventually thermalizes, $T_{\alpha \beta}(t) \rightarrow 0$, losing any memory of the initial excitation.

\begin{figure}[t]
	\includegraphics[width =\columnwidth]{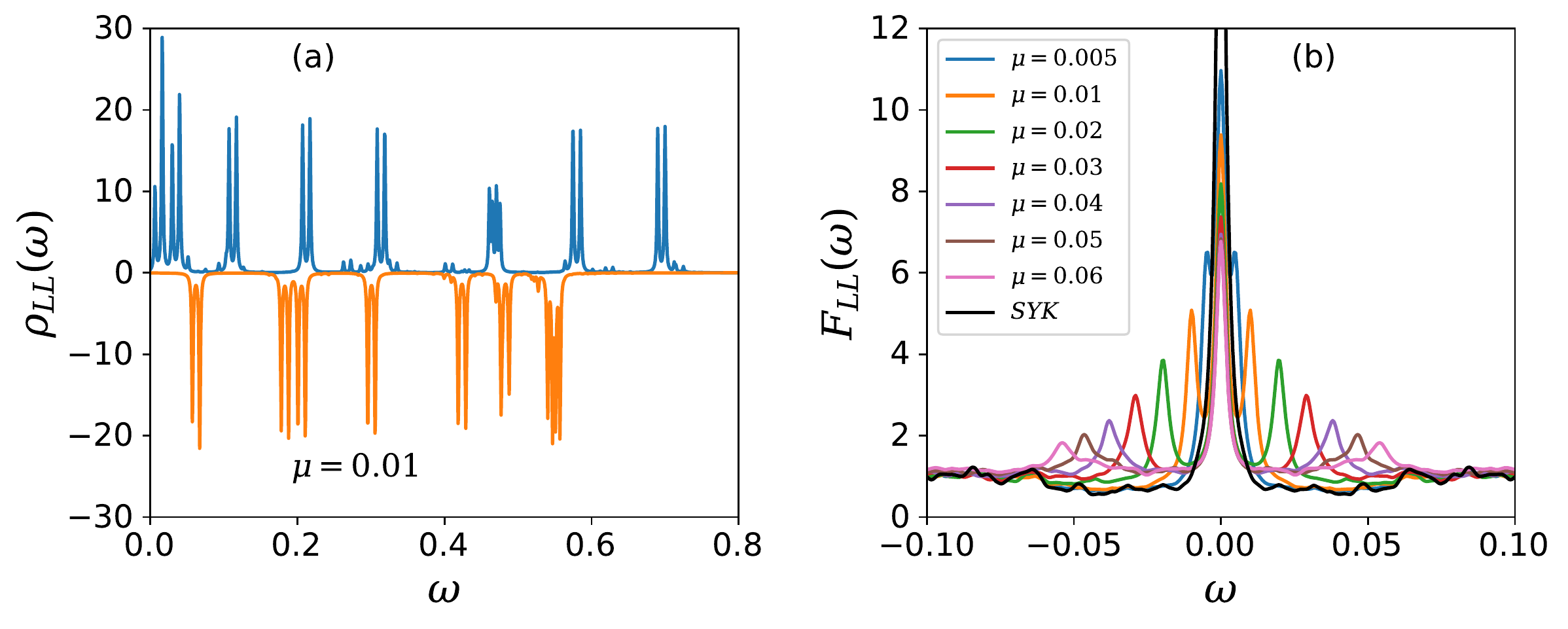}
	\caption{Spectral properties of the MQ model from exact diagonalization for $2N=16$ Majorana fermions. (a) spectral function $\rho_{LL}$ for two distinct disorder realizations, showing doublets of states split by $\omega_D$. (b) Auto-correlation $F_{LL}(\omega)$ for various $\mu$, each averaged over 50 disorder realizations.
	\label{fig:spectral_ED}}
\end{figure}

\emph{Spectrum from ED.}-- The corresponding ED results for the spectral function $\rho_{LL}(\omega)$ are shown in Fig.~\ref{fig:spectral_ED}. 
\begin{figure*}[t]
	\includegraphics[width=\textwidth]{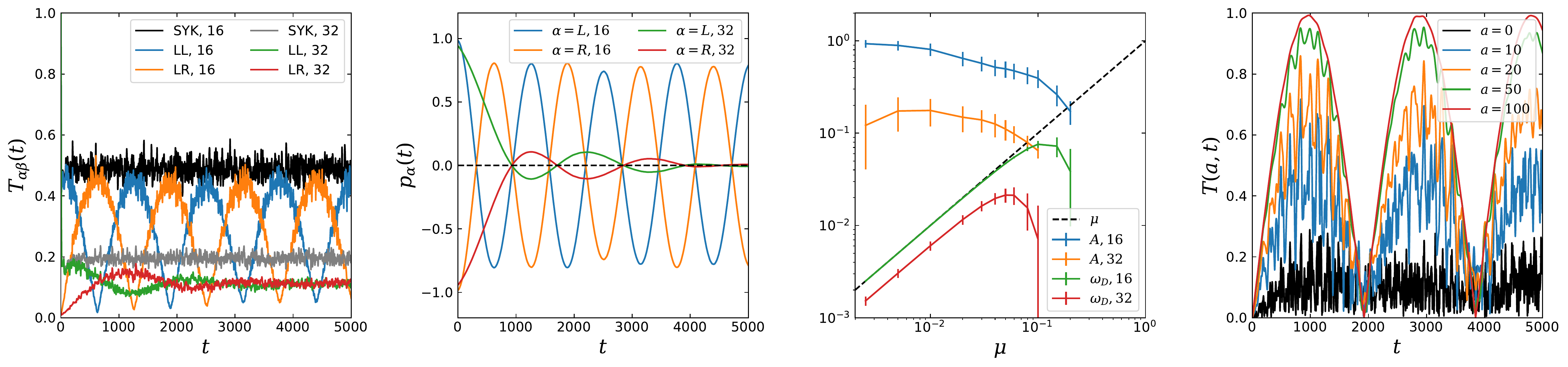}
	\caption{Revival dynamics from ED with $\mu = 0.005$. (a) Transmission amplitudes $T_{\alpha \beta}(t)$ for a single fermion mode $j$ and $2N = 16, 32$, averaged over 50 disorder realizations. The SYK result ($\mu=0$) is plotted for reference. (b) Corresponding fermion parity oscillations $p_\alpha(t)$. (c) Scaling of the oscillation frequency $\omega_D$ and amplitude $A$ of the parity oscillations in (b) as function of $\mu$. The error bars represent the standard deviation computed from 50 disorder realizations. (d) Transmission amplitude $T(a,t)$ of low-energy states $\ket{\psi^j_{R,a}}$ [Eq.~(\ref{eq:low-energy-states})], for $2N=32$ and a single disorder realization. 
	\label{fig:ED}}
\end{figure*}
We find no evidence for a conformal tower of states, presumably due to finite-$N$ effects. Instead, for $\mu \ll 1$ the spectral function consists of a collection of doublets separated by a splitting $\omega_D$. The position of these doublets in energy is uncorrelated between different disorder realizations, see Fig.~\ref{fig:spectral_ED}a. As a result, averaging the spectral function over disorder does not produce the spectral peaks expected from the large-$N$ solution. However, averaging the spectral auto-correlation function $F_{\alpha \beta}(\omega)$ [Eq.~(\ref{eq:conv})] retains a clear signal at the doublet splitting frequency $\omega_D$, while high-frequency components are washed out and roughly follow the SYK continuum, see Fig.~\ref{fig:spectral_ED}b.

\emph{Revival dynamics in ED}.-- Fig.~\ref{fig:ED} shows transmission amplitudes $T_{\alpha \beta}(t)$ for system sizes $2N=16$ and $32$~\footnote{We consider the simplest case with $N ~ \mathrm{mod} ~ 8=0$, corresponding to the Gaussian orthogonal ensemble (GOE) in the random matrix theory classification for a single SYK model. We thus avoid spectral degeneracies in the SYK spectrum, present for the unitary (GUE) and symplectic (GSE) ensembles, which lead to more complicated dynamics that obscure the revival physics under study.}
For a single disorder realization, we observe rapid oscillations modulated by a much slower envelope. The fast dynamics originate from the discreteness of the spectrum in ED ($\lambda \sim N e^{-N}$ is the typical level spacing) and vary strongly between disorder realizations. Averaging $T_{\alpha \beta}(t)$ over disorder smooths them out and retains only the envelope, see Fig.~\ref{fig:ED}a. A cleaner way to observe the revival oscillations is from the time evolution of \emph{fermionic parities} on each side, $p_\alpha(t) = \langle \Psi | P_\alpha(t) | \Psi \rangle$ with $P_\alpha = (-i)^{N/2} \prod_k \chi^k_\alpha$ and the initial excitation $\ket{ \Psi} = \chi^j_R \ket{\Psi_0}$. This observable tracks the propagation of fermion modes between the two sides regardless of the precise quantum state, and thus is less sensitive to disorder, as shown in Fig.~\ref{fig:ED}b. Note that the revival oscillations are less sharply peaked than in large--$N$, since  $F_{\alpha\beta}(\omega)$ shows a single peak at frequency $\omega_D$ rather than a series of evenly-spaced peaks. The frequency $\omega_D$ of the envelope is precisely the doublet splitting identified in Fig.~\ref{fig:spectral_ED}. It scales linearly with $\mu$ for $\mu < \lambda$ and sub-linearly for $\lambda < \mu \ll 1$, see Fig.~\ref{fig:ED}c. The latter is the regime of interest for the wormhole solution -- however, a power-law $\omega_D \sim \mu^{2/3}$, as predicted by scaling arguments~\cite{Qi2018} and obtained in SD numerics, cannot be extracted from ED.

Contrary to the expectation that ED should approach the saddle-point results as $N$ increases, the amplitude of oscillations in $T_{\alpha \beta}(t)$ and $p_\alpha(t)$ actually goes \emph{down} significantly when increasing $2N$ from 16 to 32 (see Fig.~\ref{fig:ED}a-c). This is because conformal towers of states are absent for small $N$. The observed revivals in ED do not rely on a series of evenly-spaced spectral peaks, but rather on a series of spectral doublets with roughly the same splitting, as shown in Fig.~\ref{fig:spectral_ED}. These doublets are a finite-size effect that exists only when $\mu \ll \lambda$. As such, when either $\mu$ or $N$ is increased the distribution of doublet splitting frequencies becomes broader~\cite{SuppMat}, leading to a reduced oscillation amplitude, until the doublets disappear into the SYK continuum. We also find an exact result for the averaged revivals when $2N=16$, which explains the corresponding absence of decay in Fig.~\ref{fig:ED}a-b~\cite{SuppMat}.

For the same reason, transmissions in ED are improved drastically when considering \emph{low-energy} excitations initially localized on one side. These can be generated by applying the projection
\begin{equation}
    \ket{\psi^j_{\alpha, a}} = \frac{1}{\sqrt C}\sum_n e^{-\frac{a (E_n-E_0)}{|E_0|}} \ket{n} \bra{n} \chi^j_\alpha \ket{0},
    \label{eq:low-energy-states}
\end{equation}
with normalization $C$. For $a > 0$ this shifts the spectral weight of the excitation to a few low-energy doublets of states. As a result the transmissions $T_{a}(t) = |\langle \psi^j_{L, a}(t)|\psi^j_{R, a} \rangle|$ show oscillations with an amplitude that increases with $a$ and saturates at $T_a(t_\mathrm{tr}) \sim 1$, see Fig.~\ref{fig:ED}d.

\emph{Discussion and outlook.}-- We studied dynamical properties of the Maldacena-Qi model, believed to be holographically dual to a traversable wormhole. In the large-$N$ limit, sharp revival oscillations at frequency $\sim \mu^{2/3}$ are observed in fermion transmission amplitudes, consistent with predictions of quantum gravity. The revivals arise from a conformal tower of states that emerges from the SYK continuum, and for weak couplings $\mu$ their amplitude decays as a power-law in time.
For small $N$, no towers of states are visible in exact diagonalization. Revivals instead rely on a series of approximately equally-split doublets, and the revival amplitude \emph{decreases} with $N$. Only \emph{low-energy} states localized in one system can be transmitted with probability $\sim 1$. The associated fermion parity oscillations could provide an experimental probe of revivals in the proposed platforms for physical realization~\cite{Etienne2019}, by employing parity measurements after ``severing'' the wormhole (quench to $\mu =0$).

Our work leaves several interesting open questions that are relevant to future experimental explorations of revivals in coupled SYK models. How do the dynamics of the MQ model change with increasing $N$, from being dominated by finite-size effects to arising from conformal behavior? In other words, how does the dual gravitational description emerge in a system with finite $N$? 
These questions are out of reach of exact diagonalization, but progress could be made with Krylov subspace methods, recently used to study quantum chaos in the SYK model~\cite{kobrin2020}. Other interesting avenues include understanding the role of perfectly correlated disorder in the transmission of excitations between the two subsystems, and relating the present results to circuit-based teleportation protocols using traversable wormholes~\cite{Brown2019, Gao2019}.

\emph{Note:} After completion of this work, we learned about a related parallel study by X.-L. Qi and P. Zhang that focuses on finite-temperature effects in the MQ model~\cite{QiZhang2020}.

\emph{Acknowledgements.}--
We thank O. Can, B. Kobrin, M. Rozali, X.-L. Qi, C. Li, S. Sahoo and especially J. Maldacena and A. Milekhin for useful discussions and feedback on an earlier version of this manuscript. Research was supported by SBQMI at UBC, UBC ARC Sockeye, NSERC, and CIfAR. E.L.H. acknowledges the hospitality of KITP where some of this work was completed, with support from the Heising-Simons Foundation, the Simons Foundation, and National Science Foundation Grant No. NSF PHY-1748958.

\bibliography{revival}

\begin{thebibliography}{36}%
\makeatletter
\providecommand \@ifxundefined [1]{%
 \@ifx{#1\undefined}
}%
\providecommand \@ifnum [1]{%
 \ifnum #1\expandafter \@firstoftwo
 \else \expandafter \@secondoftwo
 \fi
}%
\providecommand \@ifx [1]{%
 \ifx #1\expandafter \@firstoftwo
 \else \expandafter \@secondoftwo
 \fi
}%
\providecommand \natexlab [1]{#1}%
\providecommand \enquote  [1]{``#1''}%
\providecommand \bibnamefont  [1]{#1}%
\providecommand \bibfnamefont [1]{#1}%
\providecommand \citenamefont [1]{#1}%
\providecommand \href@noop [0]{\@secondoftwo}%
\providecommand \href [0]{\begingroup \@sanitize@url \@href}%
\providecommand \@href[1]{\@@startlink{#1}\@@href}%
\providecommand \@@href[1]{\endgroup#1\@@endlink}%
\providecommand \@sanitize@url [0]{\catcode `\\12\catcode `\$12\catcode
  `\&12\catcode `\#12\catcode `\^12\catcode `\_12\catcode `\%12\relax}%
\providecommand \@@startlink[1]{}%
\providecommand \@@endlink[0]{}%
\providecommand \url  [0]{\begingroup\@sanitize@url \@url }%
\providecommand \@url [1]{\endgroup\@href {#1}{\urlprefix }}%
\providecommand \urlprefix  [0]{URL }%
\providecommand \Eprint [0]{\href }%
\providecommand \doibase [0]{http://dx.doi.org/}%
\providecommand \selectlanguage [0]{\@gobble}%
\providecommand \bibinfo  [0]{\@secondoftwo}%
\providecommand \bibfield  [0]{\@secondoftwo}%
\providecommand \translation [1]{[#1]}%
\providecommand \BibitemOpen [0]{}%
\providecommand \bibitemStop [0]{}%
\providecommand \bibitemNoStop [0]{.\EOS\space}%
\providecommand \EOS [0]{\spacefactor3000\relax}%
\providecommand \BibitemShut  [1]{\csname bibitem#1\endcsname}%
\let\auto@bib@innerbib\@empty
\bibitem [{\citenamefont {Einstein}\ and\ \citenamefont
  {Rosen}(1935)}]{Einstein35}%
  \BibitemOpen
  \bibfield  {author} {\bibinfo {author} {\bibfnamefont {A.}~\bibnamefont
  {Einstein}}\ and\ \bibinfo {author} {\bibfnamefont {N.}~\bibnamefont
  {Rosen}},\ }\bibfield  {title} {\enquote {\bibinfo {title} {The particle
  problem in the general theory of relativity},}\ }\href {\doibase
  10.1103/PhysRev.48.73} {\bibfield  {journal} {\bibinfo  {journal} {Phys.
  Rev.}\ }\textbf {\bibinfo {volume} {48}},\ \bibinfo {pages} {73--77}
  (\bibinfo {year} {1935})}\BibitemShut {NoStop}%
\bibitem [{\citenamefont {Morris}\ and\ \citenamefont
  {Thorne}(1988)}]{Thorne88}%
  \BibitemOpen
  \bibfield  {author} {\bibinfo {author} {\bibfnamefont {Michael~S.}\
  \bibnamefont {Morris}}\ and\ \bibinfo {author} {\bibfnamefont {Kip~S.}\
  \bibnamefont {Thorne}},\ }\bibfield  {title} {\enquote {\bibinfo {title}
  {Wormholes in spacetime and their use for interstellar travel: A tool for
  teaching general relativity},}\ }\href {\doibase 10.1119/1.15620} {\bibfield
  {journal} {\bibinfo  {journal} {Am. J. Phys.}\ }\textbf {\bibinfo {volume}
  {56}},\ \bibinfo {pages} {395--412} (\bibinfo {year} {1988})}\BibitemShut
  {NoStop}%
\bibitem [{\citenamefont {{Gao}}\ \emph {et~al.}(2017)\citenamefont {{Gao}},
  \citenamefont {{Jafferis}},\ and\ \citenamefont {{Wall}}}]{Gao2017}%
  \BibitemOpen
  \bibfield  {author} {\bibinfo {author} {\bibfnamefont {P.}~\bibnamefont
  {{Gao}}}, \bibinfo {author} {\bibfnamefont {D.~L.}\ \bibnamefont
  {{Jafferis}}}, \ and\ \bibinfo {author} {\bibfnamefont {A.~C.}\ \bibnamefont
  {{Wall}}},\ }\bibfield  {title} {\enquote {\bibinfo {title} {{Traversable
  wormholes via a double trace deformation}},}\ }\href {\doibase
  10.1007/JHEP12(2017)151} {\bibfield  {journal} {\bibinfo  {journal} {J. High
  Energy Phys.}\ }\textbf {\bibinfo {volume} {12}},\ \bibinfo {eid} {151}
  (\bibinfo {year} {2017})}\BibitemShut {NoStop}%
\bibitem [{\citenamefont {Maldacena}\ \emph {et~al.}(2017)\citenamefont
  {Maldacena}, \citenamefont {Stanford},\ and\ \citenamefont
  {Yang}}]{Yang2017}%
  \BibitemOpen
  \bibfield  {author} {\bibinfo {author} {\bibfnamefont {Juan}\ \bibnamefont
  {Maldacena}}, \bibinfo {author} {\bibfnamefont {Douglas}\ \bibnamefont
  {Stanford}}, \ and\ \bibinfo {author} {\bibfnamefont {Zhenbin}\ \bibnamefont
  {Yang}},\ }\bibfield  {title} {\enquote {\bibinfo {title} {Diving into
  traversable wormholes},}\ }\href {\doibase 10.1002/prop.201700034} {\bibfield
   {journal} {\bibinfo  {journal} {Fortschr. Phys.}\ }\textbf {\bibinfo
  {volume} {65}},\ \bibinfo {pages} {1700034} (\bibinfo {year}
  {2017})}\BibitemShut {NoStop}%
\bibitem [{\citenamefont {{Maldacena}}\ \emph {et~al.}(2018)\citenamefont
  {{Maldacena}}, \citenamefont {{Milekhin}},\ and\ \citenamefont
  {{Popov}}}]{Maldacena2018}%
  \BibitemOpen
  \bibfield  {author} {\bibinfo {author} {\bibfnamefont {Juan}\ \bibnamefont
  {{Maldacena}}}, \bibinfo {author} {\bibfnamefont {Alexey}\ \bibnamefont
  {{Milekhin}}}, \ and\ \bibinfo {author} {\bibfnamefont {Fedor}\ \bibnamefont
  {{Popov}}},\ }\bibfield  {title} {\enquote {\bibinfo {title} {{Traversable
  wormholes in four dimensions}},}\ }\href
  {https://ui.adsabs.harvard.edu/abs/2018arXiv180704726M} {\bibfield  {journal}
  {\bibinfo  {journal} {arXiv:1807.04726}\ } (\bibinfo {year}
  {2018})}\BibitemShut {NoStop}%
\bibitem [{\citenamefont {Bak}\ \emph {et~al.}(2018)\citenamefont {Bak},
  \citenamefont {Kim},\ and\ \citenamefont {Yi}}]{Bak2018}%
  \BibitemOpen
  \bibfield  {author} {\bibinfo {author} {\bibfnamefont {Dongsu}\ \bibnamefont
  {Bak}}, \bibinfo {author} {\bibfnamefont {Chanju}\ \bibnamefont {Kim}}, \
  and\ \bibinfo {author} {\bibfnamefont {Sang-Heon}\ \bibnamefont {Yi}},\
  }\bibfield  {title} {\enquote {\bibinfo {title} {Bulk view of teleportation
  and traversable wormholes},}\ }\href
  {https://link.springer.com/article/10.1007/JHEP08(2018)140} {\bibfield
  {journal} {\bibinfo  {journal} {J. High Energy Phys.}\ }\textbf {\bibinfo
  {volume} {2018}},\ \bibinfo {pages} {140} (\bibinfo {year}
  {2018})}\BibitemShut {NoStop}%
\bibitem [{\citenamefont {Gao}\ and\ \citenamefont {Liu}(2019)}]{GaoLiu2019}%
  \BibitemOpen
  \bibfield  {author} {\bibinfo {author} {\bibfnamefont {Ping}\ \bibnamefont
  {Gao}}\ and\ \bibinfo {author} {\bibfnamefont {Hong}\ \bibnamefont {Liu}},\
  }\bibfield  {title} {\enquote {\bibinfo {title} {Regenesis and quantum
  traversable wormholes},}\ }\href
  {https://link.springer.com/article/10.1007%2FJHEP10%282019%29048} {\bibfield
  {journal} {\bibinfo  {journal} {J. High Energy Phys.}\ }\textbf {\bibinfo
  {volume} {2019}},\ \bibinfo {pages} {48} (\bibinfo {year}
  {2019})}\BibitemShut {NoStop}%
\bibitem [{\citenamefont {Fu}\ \emph {et~al.}(2019)\citenamefont {Fu},
  \citenamefont {Grado-White},\ and\ \citenamefont {Marolf}}]{Marolf2019}%
  \BibitemOpen
  \bibfield  {author} {\bibinfo {author} {\bibfnamefont {Zicao}\ \bibnamefont
  {Fu}}, \bibinfo {author} {\bibfnamefont {Brianna}\ \bibnamefont
  {Grado-White}}, \ and\ \bibinfo {author} {\bibfnamefont {Donald}\
  \bibnamefont {Marolf}},\ }\bibfield  {title} {\enquote {\bibinfo {title}
  {Traversable asymptotically flat wormholes with short transit times},}\
  }\href {\doibase 10.1088/1361-6382/ab56e4} {\bibfield  {journal} {\bibinfo
  {journal} {Classical and Quantum Gravity}\ }\textbf {\bibinfo {volume}
  {36}},\ \bibinfo {pages} {245018} (\bibinfo {year} {2019})}\BibitemShut
  {NoStop}%
\bibitem [{\citenamefont {Bak}\ \emph {et~al.}(2019)\citenamefont {Bak},
  \citenamefont {Kim},\ and\ \citenamefont {Yi}}]{Bak2019}%
  \BibitemOpen
  \bibfield  {author} {\bibinfo {author} {\bibfnamefont {Dongsu}\ \bibnamefont
  {Bak}}, \bibinfo {author} {\bibfnamefont {Chanju}\ \bibnamefont {Kim}}, \
  and\ \bibinfo {author} {\bibfnamefont {Sang-Heon}\ \bibnamefont {Yi}},\
  }\bibfield  {title} {\enquote {\bibinfo {title} {Experimental probes of
  traversable wormholes},}\ }\href
  {https://link.springer.com/article/10.1007/JHEP12(2019)005} {\bibfield
  {journal} {\bibinfo  {journal} {Journal of High Energy Physics}\ }\textbf
  {\bibinfo {volume} {2019}},\ \bibinfo {pages} {5} (\bibinfo {year}
  {2019})}\BibitemShut {NoStop}%
\bibitem [{\citenamefont {Sachdev}\ and\ \citenamefont {Ye}(1993)}]{SY1996}%
  \BibitemOpen
  \bibfield  {author} {\bibinfo {author} {\bibfnamefont {Subir}\ \bibnamefont
  {Sachdev}}\ and\ \bibinfo {author} {\bibfnamefont {Jinwu}\ \bibnamefont
  {Ye}},\ }\bibfield  {title} {\enquote {\bibinfo {title} {Gapless spin-fluid
  ground state in a random quantum heisenberg magnet},}\ }\href {\doibase
  10.1103/PhysRevLett.70.3339} {\bibfield  {journal} {\bibinfo  {journal}
  {Phys. Rev. Lett.}\ }\textbf {\bibinfo {volume} {70}},\ \bibinfo {pages}
  {3339} (\bibinfo {year} {1993})}\BibitemShut {NoStop}%
\bibitem [{\citenamefont {Kitaev}(2015)}]{Kitaev2015}%
  \BibitemOpen
  \bibfield  {author} {\bibinfo {author} {\bibfnamefont {A.}~\bibnamefont
  {Kitaev}},\ }\bibfield  {title} {\enquote {\bibinfo {title} {A simple model
  of quantum holography},}\ }\href
  {http://online.kitp.ucsb.edu/online/entangled15/} {\bibfield  {journal}
  {\bibinfo  {journal} {in KITP Strings Seminar and Entanglement 2015 Program}\
  } (\bibinfo {year} {2015})}\BibitemShut {NoStop}%
\bibitem [{\citenamefont {Sachdev}(2015)}]{Sachdev2015}%
  \BibitemOpen
  \bibfield  {author} {\bibinfo {author} {\bibfnamefont {Subir}\ \bibnamefont
  {Sachdev}},\ }\bibfield  {title} {\enquote {\bibinfo {title}
  {Bekenstein-hawking entropy and strange metals},}\ }\href {\doibase
  10.1103/PhysRevX.5.041025} {\bibfield  {journal} {\bibinfo  {journal} {Phys.
  Rev. X}\ }\textbf {\bibinfo {volume} {5}},\ \bibinfo {pages} {041025}
  (\bibinfo {year} {2015})}\BibitemShut {NoStop}%
\bibitem [{\citenamefont {Maldacena}\ and\ \citenamefont
  {Stanford}(2016)}]{Maldacena2016}%
  \BibitemOpen
  \bibfield  {author} {\bibinfo {author} {\bibfnamefont {Juan}\ \bibnamefont
  {Maldacena}}\ and\ \bibinfo {author} {\bibfnamefont {Douglas}\ \bibnamefont
  {Stanford}},\ }\bibfield  {title} {\enquote {\bibinfo {title} {Remarks on the
  sachdev-ye-kitaev model},}\ }\href {\doibase 10.1103/PhysRevD.94.106002}
  {\bibfield  {journal} {\bibinfo  {journal} {Phys. Rev. D}\ }\textbf {\bibinfo
  {volume} {94}},\ \bibinfo {pages} {106002} (\bibinfo {year}
  {2016})}\BibitemShut {NoStop}%
\bibitem [{\citenamefont {Maldacena}\ \emph {et~al.}(2016)\citenamefont
  {Maldacena}, \citenamefont {Shenker},\ and\ \citenamefont
  {Stanford}}]{Maldacena2016b}%
  \BibitemOpen
  \bibfield  {author} {\bibinfo {author} {\bibfnamefont {Juan}\ \bibnamefont
  {Maldacena}}, \bibinfo {author} {\bibfnamefont {Stephen~H.}\ \bibnamefont
  {Shenker}}, \ and\ \bibinfo {author} {\bibfnamefont {Douglas}\ \bibnamefont
  {Stanford}},\ }\bibfield  {title} {\enquote {\bibinfo {title} {A bound on
  chaos},}\ }\href {\doibase 10.1007/JHEP08(2016)106} {\bibfield  {journal}
  {\bibinfo  {journal} {J. High Energy Phys.}\ }\textbf {\bibinfo {volume}
  {2016}},\ \bibinfo {pages} {106} (\bibinfo {year} {2016})}\BibitemShut
  {NoStop}%
\bibitem [{\citenamefont {Danshita}\ \emph {et~al.}(2017)\citenamefont
  {Danshita}, \citenamefont {Hanada},\ and\ \citenamefont
  {Tezuka}}]{Danshita2017}%
  \BibitemOpen
  \bibfield  {author} {\bibinfo {author} {\bibfnamefont {Ippei}\ \bibnamefont
  {Danshita}}, \bibinfo {author} {\bibfnamefont {Masanori}\ \bibnamefont
  {Hanada}}, \ and\ \bibinfo {author} {\bibfnamefont {Masaki}\ \bibnamefont
  {Tezuka}},\ }\bibfield  {title} {\enquote {\bibinfo {title} {Creating and
  probing the sachdev–ye–kitaev model with ultracold gases: Towards
  experimental studies of quantum gravity},}\ }\href {\doibase
  10.1093/ptep/ptx108} {\bibfield  {journal} {\bibinfo  {journal} {Prog. Theor.
  Exp. Phys.}\ }\textbf {\bibinfo {volume} {2017}},\ \bibinfo {pages} {083I01}
  (\bibinfo {year} {2017})}\BibitemShut {NoStop}%
\bibitem [{\citenamefont {Pikulin}\ and\ \citenamefont
  {Franz}(2017)}]{Pikulin2017}%
  \BibitemOpen
  \bibfield  {author} {\bibinfo {author} {\bibfnamefont {D.~I.}\ \bibnamefont
  {Pikulin}}\ and\ \bibinfo {author} {\bibfnamefont {M.}~\bibnamefont
  {Franz}},\ }\bibfield  {title} {\enquote {\bibinfo {title} {Black hole on a
  chip: Proposal for a physical realization of the sachdev-ye-kitaev model in a
  solid-state system},}\ }\href {\doibase 10.1103/PhysRevX.7.031006} {\bibfield
   {journal} {\bibinfo  {journal} {Phys. Rev. X}\ }\textbf {\bibinfo {volume}
  {7}},\ \bibinfo {pages} {031006} (\bibinfo {year} {2017})}\BibitemShut
  {NoStop}%
\bibitem [{\citenamefont {Chew}\ \emph {et~al.}(2017)\citenamefont {Chew},
  \citenamefont {Essin},\ and\ \citenamefont {Alicea}}]{Alicea2017}%
  \BibitemOpen
  \bibfield  {author} {\bibinfo {author} {\bibfnamefont {Aaron}\ \bibnamefont
  {Chew}}, \bibinfo {author} {\bibfnamefont {Andrew}\ \bibnamefont {Essin}}, \
  and\ \bibinfo {author} {\bibfnamefont {Jason}\ \bibnamefont {Alicea}},\
  }\bibfield  {title} {\enquote {\bibinfo {title} {Approximating the
  sachdev-ye-kitaev model with majorana wires},}\ }\href {\doibase
  10.1103/PhysRevB.96.121119} {\bibfield  {journal} {\bibinfo  {journal} {Phys.
  Rev. B}\ }\textbf {\bibinfo {volume} {96}},\ \bibinfo {pages} {121119(R)}
  (\bibinfo {year} {2017})}\BibitemShut {NoStop}%
\bibitem [{\citenamefont {Chen}\ \emph {et~al.}(2018)\citenamefont {Chen},
  \citenamefont {Ilan}, \citenamefont {de~Juan}, \citenamefont {Pikulin},\ and\
  \citenamefont {Franz}}]{achen2018}%
  \BibitemOpen
  \bibfield  {author} {\bibinfo {author} {\bibfnamefont {Anffany}\ \bibnamefont
  {Chen}}, \bibinfo {author} {\bibfnamefont {R.}~\bibnamefont {Ilan}}, \bibinfo
  {author} {\bibfnamefont {F.}~\bibnamefont {de~Juan}}, \bibinfo {author}
  {\bibfnamefont {D.~I.}\ \bibnamefont {Pikulin}}, \ and\ \bibinfo {author}
  {\bibfnamefont {M.}~\bibnamefont {Franz}},\ }\bibfield  {title} {\enquote
  {\bibinfo {title} {Quantum holography in a graphene flake with an irregular
  boundary},}\ }\href {\doibase 10.1103/PhysRevLett.121.036403} {\bibfield
  {journal} {\bibinfo  {journal} {Phys. Rev. Lett.}\ }\textbf {\bibinfo
  {volume} {121}},\ \bibinfo {pages} {036403} (\bibinfo {year}
  {2018})}\BibitemShut {NoStop}%
\bibitem [{\citenamefont {Franz}\ and\ \citenamefont
  {Rozali}(2018)}]{Rozali2018}%
  \BibitemOpen
  \bibfield  {author} {\bibinfo {author} {\bibfnamefont {Marcel}\ \bibnamefont
  {Franz}}\ and\ \bibinfo {author} {\bibfnamefont {Moshe}\ \bibnamefont
  {Rozali}},\ }\bibfield  {title} {\enquote {\bibinfo {title} {Mimicking black
  hole event horizons in atomic and solid-state systems},}\ }\href {\doibase
  10.1038/s41578-018-0058-z} {\bibfield  {journal} {\bibinfo  {journal} {Nat.
  Rev. Mater.}\ }\textbf {\bibinfo {volume} {3}},\ \bibinfo {pages} {491}
  (\bibinfo {year} {2018})}\BibitemShut {NoStop}%
\bibitem [{\citenamefont {Altland}\ \emph {et~al.}(2019)\citenamefont
  {Altland}, \citenamefont {Bagrets},\ and\ \citenamefont
  {Kamenev}}]{Altland2019}%
  \BibitemOpen
  \bibfield  {author} {\bibinfo {author} {\bibfnamefont {Alexander}\
  \bibnamefont {Altland}}, \bibinfo {author} {\bibfnamefont {Dmitry}\
  \bibnamefont {Bagrets}}, \ and\ \bibinfo {author} {\bibfnamefont {Alex}\
  \bibnamefont {Kamenev}},\ }\bibfield  {title} {\enquote {\bibinfo {title}
  {Sachdev-ye-kitaev non-fermi-liquid correlations in nanoscopic quantum
  transport},}\ }\href {\doibase 10.1103/PhysRevLett.123.226801} {\bibfield
  {journal} {\bibinfo  {journal} {Phys. Rev. Lett.}\ }\textbf {\bibinfo
  {volume} {123}},\ \bibinfo {pages} {226801} (\bibinfo {year}
  {2019})}\BibitemShut {NoStop}%
\bibitem [{\citenamefont {{Maldacena}}\ and\ \citenamefont
  {{Qi}}(2018)}]{Qi2018}%
  \BibitemOpen
  \bibfield  {author} {\bibinfo {author} {\bibfnamefont {Juan}\ \bibnamefont
  {{Maldacena}}}\ and\ \bibinfo {author} {\bibfnamefont {Xiao-Liang}\
  \bibnamefont {{Qi}}},\ }\bibfield  {title} {\enquote {\bibinfo {title}
  {{Eternal traversable wormhole}},}\ }\href
  {https://ui.adsabs.harvard.edu/abs/2018arXiv180400491M} {\bibfield  {journal}
  {\bibinfo  {journal} {arXiv:1804.00491}\ } (\bibinfo {year}
  {2018})}\BibitemShut {NoStop}%
\bibitem [{\citenamefont {Garc\'{\i}a-Garc\'{\i}a}\ \emph
  {et~al.}(2019)\citenamefont {Garc\'{\i}a-Garc\'{\i}a}, \citenamefont
  {Nosaka}, \citenamefont {Rosa},\ and\ \citenamefont
  {Verbaarschot}}]{Garcia2019}%
  \BibitemOpen
  \bibfield  {author} {\bibinfo {author} {\bibfnamefont {Antonio~M.}\
  \bibnamefont {Garc\'{\i}a-Garc\'{\i}a}}, \bibinfo {author} {\bibfnamefont
  {Tomoki}\ \bibnamefont {Nosaka}}, \bibinfo {author} {\bibfnamefont {Dario}\
  \bibnamefont {Rosa}}, \ and\ \bibinfo {author} {\bibfnamefont {Jacobus
  J.~M.}\ \bibnamefont {Verbaarschot}},\ }\bibfield  {title} {\enquote
  {\bibinfo {title} {Quantum chaos transition in a two-site sachdev-ye-kitaev
  model dual to an eternal traversable wormhole},}\ }\href {\doibase
  10.1103/PhysRevD.100.026002} {\bibfield  {journal} {\bibinfo  {journal}
  {Phys. Rev. D}\ }\textbf {\bibinfo {volume} {100}},\ \bibinfo {pages}
  {026002} (\bibinfo {year} {2019})}\BibitemShut {NoStop}%
\bibitem [{\citenamefont {{Maldacena}}\ and\ \citenamefont
  {{Milekhin}}()}]{Maldacena2019}%
  \BibitemOpen
  \bibfield  {author} {\bibinfo {author} {\bibfnamefont {Juan}\ \bibnamefont
  {{Maldacena}}}\ and\ \bibinfo {author} {\bibfnamefont {Alexey}\ \bibnamefont
  {{Milekhin}}},\ }\bibfield  {title} {\enquote {\bibinfo {title} {{SYK
  wormhole formation in real time}},}\ }\href
  {https://ui.adsabs.harvard.edu/abs/2019arXiv191203276M} {\bibinfo  {journal}
  {arXiv:1912.03276}\ }\BibitemShut {NoStop}%
\bibitem [{\citenamefont {Chen}\ and\ \citenamefont {Zhang}(2019)}]{Chen2019}%
  \BibitemOpen
\bibfield  {journal} {  }\bibfield  {author} {\bibinfo {author} {\bibfnamefont
  {Yiming}\ \bibnamefont {Chen}}\ and\ \bibinfo {author} {\bibfnamefont
  {Pengfei}\ \bibnamefont {Zhang}},\ }\bibfield  {title} {\enquote {\bibinfo
  {title} {Entanglement entropy of two coupled syk models and eternal
  traversable wormhole},}\ }\href
  {https://link.springer.com/article/10.1007/JHEP07(2019)033} {\bibfield
  {journal} {\bibinfo  {journal} {J. High Energy Phys.}\ }\textbf {\bibinfo
  {volume} {2019}},\ \bibinfo {pages} {33} (\bibinfo {year}
  {2019})}\BibitemShut {NoStop}%
\bibitem [{\citenamefont {{Alet}}\ \emph {et~al.}()\citenamefont {{Alet}},
  \citenamefont {{Hanada}}, \citenamefont {{Jevicki}},\ and\ \citenamefont
  {{Peng}}}]{Alet2020}%
  \BibitemOpen
  \bibfield  {author} {\bibinfo {author} {\bibfnamefont {Fabien}\ \bibnamefont
  {{Alet}}}, \bibinfo {author} {\bibfnamefont {Masanori}\ \bibnamefont
  {{Hanada}}}, \bibinfo {author} {\bibfnamefont {Antal}\ \bibnamefont
  {{Jevicki}}}, \ and\ \bibinfo {author} {\bibfnamefont {Cheng}\ \bibnamefont
  {{Peng}}},\ }\bibfield  {title} {\enquote {\bibinfo {title} {{Entanglement
  and Confinement in Coupled Quantum Systems}},}\ }\href
  {https://ui.adsabs.harvard.edu/abs/2020arXiv200103158A} {\bibinfo  {journal}
  {arXiv:2001.03158}\ }\BibitemShut {NoStop}%
\bibitem [{\citenamefont {Lantagne-Hurtubise}\ \emph
  {et~al.}(2020)\citenamefont {Lantagne-Hurtubise}, \citenamefont {Plugge},
  \citenamefont {Can},\ and\ \citenamefont {Franz}}]{Etienne2019}%
  \BibitemOpen
\bibfield  {journal} {  }\bibfield  {author} {\bibinfo {author} {\bibfnamefont
  {\'Etienne}\ \bibnamefont {Lantagne-Hurtubise}}, \bibinfo {author}
  {\bibfnamefont {Stephan}\ \bibnamefont {Plugge}}, \bibinfo {author}
  {\bibfnamefont {Oguzhan}\ \bibnamefont {Can}}, \ and\ \bibinfo {author}
  {\bibfnamefont {Marcel}\ \bibnamefont {Franz}},\ }\bibfield  {title}
  {\enquote {\bibinfo {title} {Diagnosing quantum chaos in many-body systems
  using entanglement as a resource},}\ }\href {\doibase
  10.1103/PhysRevResearch.2.013254} {\bibfield  {journal} {\bibinfo  {journal}
  {Phys. Rev. Research}\ }\textbf {\bibinfo {volume} {2}},\ \bibinfo {pages}
  {013254} (\bibinfo {year} {2020})}\BibitemShut {NoStop}%
\bibitem [{Sup()}]{SuppMat}%
  \BibitemOpen
  \href@noop {} {}\bibinfo {note} {For further discussion, additional data and
  analysis, see the supplementary material available at [online
  link].}\BibitemShut {Stop}%
\bibitem [{Note1()}]{Note1}%
  \BibitemOpen
  \bibinfo {note} {In order to observe revivals, it is crucial to extract more
  than just an energy gap from saddle-point equations or exact diagonalization.
  A single-scale expression $G_{LL}(\tau ) \sim e^{-E_\protect \mathrm
  {gap}\tau } \to G_{LL}(t) \sim e^{-iE_\protect \mathrm {gap}t}$~\cite
  {Qi2018,Garcia2019} does \protect \emph {not} predict revivals, cf.
  Eqs.~\protect \textup {\hbox {\mathsurround \z@ \protect \normalfont
  (\ignorespaces \ref {m7}\unskip \@@italiccorr )}}. In imaginary-time
  formulation, revivals associated with higher energy scales (conformal tower)
  are hidden in the early-time, non-exponential transient behavior of
  $G_{LL}(\tau )$.}\BibitemShut {Stop}%
\bibitem [{Note2()}]{Note2}%
  \BibitemOpen
  \bibinfo {note} {The boundary graviton states are expected to occur at higher
  order in the $1/N$ expansion, and thus do not appear in our saddle-point
  calculations. We thank J. Maldacena and A. Milekhin for pointing this out to
  us.}\BibitemShut {Stop}%
\bibitem [{\citenamefont {Qi}\ and\ \citenamefont {Zhang}(2020)}]{QiZhang2020}%
  \BibitemOpen
  \bibfield  {author} {\bibinfo {author} {\bibfnamefont {Xiao-Liang}\
  \bibnamefont {Qi}}\ and\ \bibinfo {author} {\bibfnamefont {Pengfei}\
  \bibnamefont {Zhang}},\ }\bibfield  {title} {\enquote {\bibinfo {title} {The
  coupled syk model at finite temperature},}\ }\href@noop {} {\bibfield
  {journal} {\bibinfo  {journal} {arXiv preprint arXiv:2003.03916}\ } (\bibinfo
  {year} {2020})}\BibitemShut {NoStop}%
\bibitem [{Note3()}]{Note3}%
  \BibitemOpen
  \bibinfo {note} {We consider the simplest case with $N ~ \protect \mathrm
  {mod} ~ 8=0$, corresponding to the Gaussian orthogonal ensemble (GOE) in the
  random matrix theory classification for a single SYK model. We thus avoid
  spectral degeneracies in the SYK spectrum, present for the unitary (GUE) and
  symplectic (GSE) ensembles, which lead to more complicated dynamics that
  obscure the revival physics under study.}\BibitemShut {Stop}%
\bibitem [{\citenamefont {Kobrin}\ \emph {et~al.}(2020)\citenamefont {Kobrin},
  \citenamefont {Yang}, \citenamefont {Kahanamoku-Meyer}, \citenamefont
  {Olund}, \citenamefont {Moore}, \citenamefont {Stanford},\ and\ \citenamefont
  {Yao}}]{kobrin2020}%
  \BibitemOpen
  \bibfield  {author} {\bibinfo {author} {\bibfnamefont {Bryce}\ \bibnamefont
  {Kobrin}}, \bibinfo {author} {\bibfnamefont {Zhenbin}\ \bibnamefont {Yang}},
  \bibinfo {author} {\bibfnamefont {Gregory~D}\ \bibnamefont
  {Kahanamoku-Meyer}}, \bibinfo {author} {\bibfnamefont {Christopher~T}\
  \bibnamefont {Olund}}, \bibinfo {author} {\bibfnamefont {Joel~E}\
  \bibnamefont {Moore}}, \bibinfo {author} {\bibfnamefont {Douglas}\
  \bibnamefont {Stanford}}, \ and\ \bibinfo {author} {\bibfnamefont {Norman~Y}\
  \bibnamefont {Yao}},\ }\bibfield  {title} {\enquote {\bibinfo {title}
  {Many-body chaos in the sachdev-ye-kitaev model},}\ }\href
  {https://arxiv.org/abs/2002.05725} {\bibfield  {journal} {\bibinfo  {journal}
  {arXiv:2002.05725}\ } (\bibinfo {year} {2020})}\BibitemShut {NoStop}%
\bibitem [{\citenamefont {{Brown}}\ \emph {et~al.}(2019)\citenamefont
  {{Brown}}, \citenamefont {{Gharibyan}}, \citenamefont {{Leichenauer}},
  \citenamefont {{Lin}}, \citenamefont {{Nezami}}, \citenamefont {{Salton}},
  \citenamefont {{Susskind}}, \citenamefont {{Swingle}},\ and\ \citenamefont
  {{Walter}}}]{Brown2019}%
  \BibitemOpen
  \bibfield  {author} {\bibinfo {author} {\bibfnamefont {Adam~R.}\ \bibnamefont
  {{Brown}}}, \bibinfo {author} {\bibfnamefont {Hrant}\ \bibnamefont
  {{Gharibyan}}}, \bibinfo {author} {\bibfnamefont {Stefan}\ \bibnamefont
  {{Leichenauer}}}, \bibinfo {author} {\bibfnamefont {Henry~W.}\ \bibnamefont
  {{Lin}}}, \bibinfo {author} {\bibfnamefont {Sepehr}\ \bibnamefont
  {{Nezami}}}, \bibinfo {author} {\bibfnamefont {Grant}\ \bibnamefont
  {{Salton}}}, \bibinfo {author} {\bibfnamefont {Leonard}\ \bibnamefont
  {{Susskind}}}, \bibinfo {author} {\bibfnamefont {Brian}\ \bibnamefont
  {{Swingle}}}, \ and\ \bibinfo {author} {\bibfnamefont {Michael}\ \bibnamefont
  {{Walter}}},\ }\bibfield  {title} {\enquote {\bibinfo {title} {{Quantum
  Gravity in the Lab: Teleportation by Size and Traversable Wormholes}},}\
  }\href {https://ui.adsabs.harvard.edu/abs/2019arXiv191106314B} {\bibfield
  {journal} {\bibinfo  {journal} {arXiv:2002.05725}\ } (\bibinfo {year}
  {2019})}\BibitemShut {NoStop}%
\bibitem [{\citenamefont {{Gao}}\ and\ \citenamefont
  {{Jafferis}}(2019)}]{Gao2019}%
  \BibitemOpen
  \bibfield  {author} {\bibinfo {author} {\bibfnamefont {Ping}\ \bibnamefont
  {{Gao}}}\ and\ \bibinfo {author} {\bibfnamefont {Daniel~Louis}\ \bibnamefont
  {{Jafferis}}},\ }\bibfield  {title} {\enquote {\bibinfo {title} {{A
  Traversable Wormhole Teleportation Protocol in the SYK Model}},}\ }\href
  {https://ui.adsabs.harvard.edu/abs/2019arXiv191107416G} {\bibfield  {journal}
  {\bibinfo  {journal} {arXiv:1911.07416}\ } (\bibinfo {year}
  {2019})}\BibitemShut {NoStop}%
\bibitem [{\citenamefont {Banerjee}\ and\ \citenamefont
  {Altman}(2017)}]{Altman2016}%
  \BibitemOpen
  \bibfield  {author} {\bibinfo {author} {\bibfnamefont {Sumilan}\ \bibnamefont
  {Banerjee}}\ and\ \bibinfo {author} {\bibfnamefont {Ehud}\ \bibnamefont
  {Altman}},\ }\bibfield  {title} {\enquote {\bibinfo {title} {Solvable model
  for a dynamical quantum phase transition from fast to slow scrambling},}\
  }\href {\doibase 10.1103/PhysRevB.95.134302} {\bibfield  {journal} {\bibinfo
  {journal} {Phys. Rev. B}\ }\textbf {\bibinfo {volume} {95}},\ \bibinfo
  {pages} {134302} (\bibinfo {year} {2017})}\BibitemShut {NoStop}%
\bibitem [{Note4()}]{Note4}%
  \BibitemOpen
  \bibinfo {note} {We thank J. Maldacena and A. Milekhin for clarifying to us
  that this dissipative mechanism can be included in the gravitational
  description of the MQ model.}\BibitemShut {Stop}%
\end{thebibliography}%

\clearpage

\setcounter{equation}{0}
\setcounter{figure}{0}
\setcounter{table}{0}
\setcounter{page}{1}
\renewcommand{\theequation}{S\arabic{equation}}
\renewcommand{\thefigure}{S\arabic{figure}}
\renewcommand{\bibnumfmt}[1]{[S#1]}
\renewcommand{\citenumfont}[1]{S#1}

\section{Supplementary material}

\subsection{Large-$N$ saddle point equations}

The derivation of the large-$N$ saddle point equations for the problem
defined by Eqs.~(1)-(2) in the main text proceeds along
the same lines as for the original SYK model
\cite{Kitaev2015,Maldacena2016}, so we only outline the key steps
here and focus on the novel features that arise from having two coupled SYK
models. The first step is to reformulate the theory using an
imaginary-time path integral with Majorana fields represented as
anticommuting Grassmann variables. Averaging over the quenched disorder in
coupling constants $J_{ijkl}$ then leads to an action for the averaged
propagators  $G_{\alpha\beta}(\tau_1,\tau_2)$ defined by Eq.~(3). The resulting  action is quadratic in the fermion fields
and, when these are integrated out, has the following form \cite{Qi2018,Etienne2019}
\begin{widetext}
	\begin{equation}\label{app1}
		S=-N\ln{\rm Pf} (\delta_{\alpha\beta}\partial_\tau-\Sigma_{\alpha\beta})
		+{N\over 2}\int_{\tau_1,\tau_2}\sum_{\alpha,\beta}\left[\left(
		\Sigma_{\alpha\beta}(\tau_1,\tau_2)-\mu\sigma^y_{\alpha\beta}\delta(\tau_1-\tau_2)\right)
		G_{\alpha\beta}(\tau_1,\tau_2)
		-{J^2\over 4} G_{\alpha\beta}(\tau_1,\tau_2)^4\right] ,
	\end{equation}
\end{widetext}
where $\sigma^y$ is a Pauli matrix acting in the $LR$ space. The Lagrange
multiplier $\Sigma_{\alpha\beta}$ has been introduced to enforce  Eq.~(3) and can be interpreted as the
fermion self-energy. For $\mu=0$ the action \eqref{app1} describes two
decoupled SYK models \cite{Kitaev2015,Maldacena2016}.
The large-$N$ saddle-point equations are obtained by
varying the action with respect to  $\Sigma_{\alpha\beta}$ and
$G_{\alpha\beta}$, 
\begin{eqnarray}
	G_{\alpha\beta}(i\omega_n)&=&\left[i\omega_n-\mu\sigma^y-\Sigma(i\omega_n)\right]^{-1}|_{\alpha\beta}, \label{app2}
	\\
	\Sigma_{\alpha\beta}(\tau)&=&J^2 G_{\alpha\beta} (\tau)^3,\label{app3}
\end{eqnarray}
and $\omega_n=\pi T(2n+1)$ is the $n$th Matsubara frequency.

The solution of the saddle-point equations can be considerably
simplified by  exploiting symmetries. First, we notice that the MQ
Hamiltonian is invariant under the exchange $\chi_L^j\to\chi_R^j$ and
$\chi_R^j\to-\chi_L^j$ which implies $G_{LL}(\tau)=G_{RR}(\tau)$ and
$G_{LR}(\tau)=-G_{RL}(\tau)$. The full matrix propagator can be
therefore expressed in terms of its two independent components as
\begin{equation}\label{app4}
	G(\tau)=G_{LL}(\tau) +i\sigma^y G_{LR}(\tau).
\end{equation}
To further simplify the solution it is useful to rotate into a new basis in the $LR$ space  by defining
\begin{eqnarray} \label{app5}
	G_\pm(\tau)&=&G_{LL}(\tau) \pm i G_{LR}(\tau), \\
	\Sigma_\pm(\tau)&=&\Sigma_{LL}(\tau) \pm i \Sigma_{LR}(\tau). \nonumber
\end{eqnarray}
In the absence of interactions, that is when $J=0$, this rotation
diagonalizes Eq.\ \eqref{app2} which is then solved by $G^0_\pm(i\omega_n)=1/(i\omega_n\mp \mu)$. Fourier transforming into
the imaginary time domain one immediately obtains the second useful
symmetry property
\begin{equation}\label{app6}
	G^0_\mp(\tau)=-G^0_{\pm}(-\tau).
\end{equation}

In the new basis, Eq.\
\eqref{app3} can be rewritten as
\begin{equation}\label{app7}
	\Sigma_\pm(\tau)={J^2\over 4}\left[3G_\pm^2(\tau)G_\mp(\tau)+
	G_{\mp}^3(\tau)\right].
\end{equation}
If we assume for a moment that symmetry~\eqref{app6} is the property of the full interacting propagator then Eq.~\eqref{app7} implies the same symmetry for the self energy, i.e.\  $\Sigma_\mp(\tau)=-\Sigma_{\pm}(-\tau)$.
Therefore, we conclude that the symmetry \eqref{app6} of the non-interacting system is respected by the saddle-point equations in the presence of interactions. We can then substitute $G_\mp(\tau)=-G_{\pm}(-\tau)$ in Eq.\ \eqref{app7}  which makes the equations for the $+$ and $-$  channel decouple, and leads to Eq.~(4) of the main text.

\subsubsection{Real-time and -frequency formulation of SD equations}

To access transmission amplitudes and spectral features in real time and frequency, we analytically continue the final set of Schwinger-Dyson (SD) equations, Eq.~(4) of the main text, following the procedures outlined in Refs.~\cite{Altman2016,Etienne2019}.
For completeness and to keep this work self-contained, we here outline the essential steps leading to the final set of real-time SD equations that we then solve numerically.
Analytic continuation of Eq.~(4) gives
\begin{equation}
	G_+^\mathrm{ret}(\omega) =\left[\omega+i\eta-\mu-\Sigma_+^\mathrm{ret}(\omega)\right]^{-1},
	\nonumber
\end{equation}
with bare GF $G_+^{0}(\omega) = 1/(\omega+i\eta-\mu)$ giving a Lorentzian peak ($\delta$-function) in the spectral function at $\omega = \mu$.
A tractable route to obtaining the retarded self-energy $\Sigma_+^\mathrm{ret}(\omega)$ consists of Fourier transforming the imaginary-time self-energy, going over to spectral representation of Matsubara Greens functions, and finally continuing to real frequency, using steps and identities outlined in Refs.~\cite{Altman2016,Etienne2019}.
This can either be done in $LL$-$LR$ basis, afterwards adding up contributions according to Eq.~\eqref{app5}, or by directly operating on the decoupled SD equations for $\pm$ channels, given by Eq.~(4).
In both cases, for representation purposes, it is useful to retain the occupation functions of the $LL,LR$ channels given by~\cite{Etienne2019}
\begin{equation}\label{eq:app-occupations}
	n_{LL/LR}(t) = \int_{-\infty}^\infty d\omega \rho_{LL/LR}(\omega) n_F(\omega)e^{-i\omega t}~,
\end{equation}
comprising even and odd frequency-parity combinations $\rho_{LL/LR}(\omega) = \frac12[\rho_+(\omega)\pm\rho_+(-\omega)]$ of the spectral function $\rho_+(\omega) = -\frac{1}{\pi}\mathrm{Im}G_+^\mathrm{ret}(\omega)$, cf. Eq.~\eqref{app5}, and the Fermi function $n_F(\omega)$.
The latter arises after carrying out Matsubara sums that transform imaginary to real frequencies in the spectral representation of GFs.
In doing this we use various identities~\cite{Etienne2019} to relate Fermi/Bose occupation functions on frequency differences to products of individual Fermi functions, and finally find the retarded self-energy
\begin{eqnarray}\label{eq:app-selfenergy}
	\Sigma^\mathrm{ret}_+(\omega) =
	&-&2iJ^2 \int_0^\infty dt  e^{i(\omega + i\eta) t} \times \\
	&&
	\left[\mathrm{Re}[n_{LL}^3(t)] - i\mathrm{Im}[n_{LR}^3(t)]\right]~.~
	\notag
\end{eqnarray}
From here, to find a solution of the SD equations, we perform weighted numerical iterations \cite{Maldacena2016,Qi2018,Etienne2019} transforming between time and frequency representations to evaluate self-energies and Dyson equations. We declare convergence to a physical solution once propagators and spectral functions stop changing within the accessible numerical resolution.
For the results presented in this work, we employ a frequency cutoff $\omega_\mathrm{max} = 2^{11}$ with $N_\mathrm{grid} = 2^{28}-2^{29}$ grid points for Fourier transforms and the SD iteration. For the smallest $\mu$ we consider, this leads to frequency spacings $\delta\omega \lesssim \frac{1}{10} T \lesssim \frac{1}{100} \mu$.

\subsubsection{Generalization to order-$q$ SYK-interactions}

Similar simplifications as above are possible for generalized interactions with $q > 4$ SYK fermions \cite{Maldacena2016,Qi2018} in Eq.~(2) of the main text. Here we find self-energies
\begin{equation}\label{app8}
	\Sigma_{\alpha\beta}^q(\tau) = J^2 G_{\alpha\beta} (\tau)^{q-1},
\end{equation}
in place of Eq.~\eqref{app3}. The SD equations can be simplified in an analogous manner. We again switch to combinations $G_\pm(\tau)$ and find by the same steps as above
\begin{eqnarray}\label{app9}
	\Sigma_+^q(\tau) = \frac{J^2}{2^{q-1}}
	\lbrace [G_+(\tau) &-& G_+(-\tau)]^{q-1} \\
	&-& [G_+(\tau) + G_+(-\tau)]^{q-1} \rbrace.\quad
	\notag
\end{eqnarray}
This is the generalization of Eqs.~(4),\eqref{app7} to arbitrary $q$.
With regards to the analytic continuation to real time and frequency, while intermediate steps are somewhat more tedious for general $q$ (more Matsubara sums and Bose/Fermi function identities), by analogous steps one finds for the retarded self-energy
\begin{eqnarray}\label{app10}
	\Sigma^{\mathrm{ret},q}_+(\omega) =
	&-&2iJ^2 \int_0^\infty dt  e^{i(\omega + i\eta) t} \times \\
	&&
	\left[\mathrm{Re}[n_{LL}^{q-1}(t)] - i(-1)^{q/2} \mathrm{Im}[n_{LR}^{q-1}(t)]\right]~.~
	\notag
\end{eqnarray}
Here we again used $n_{LL/LR} = \frac12(n_+\pm n_-)$ with time-dependent occupation $n_\pm(t) = \int d\omega \rho_+(\pm\omega)n_F(\omega) e^{-i\omega t}$, cf. Eq.~\eqref{eq:app-occupations}.
This result allows to solve real time and frequency SD equations for the MQ model at arbitrary $q$ in a unified manner.
However note that by virtue of Eq.~\eqref{app10}, larger $q$ leads to stronger non-linearity of the self-energy, implying more difficulty in ensuring numerical stability and convergence. Indeed upon numerically implementing the above equations for $q=6$ or $q=8$, by naive comparison to the simplest $q=4$ case, we found the quality of solutions to deteriorate rapidly.

\subsection{Link between revival dynamics and spectral features}

\begin{figure}[t]
	\includegraphics[width=1.0\columnwidth]{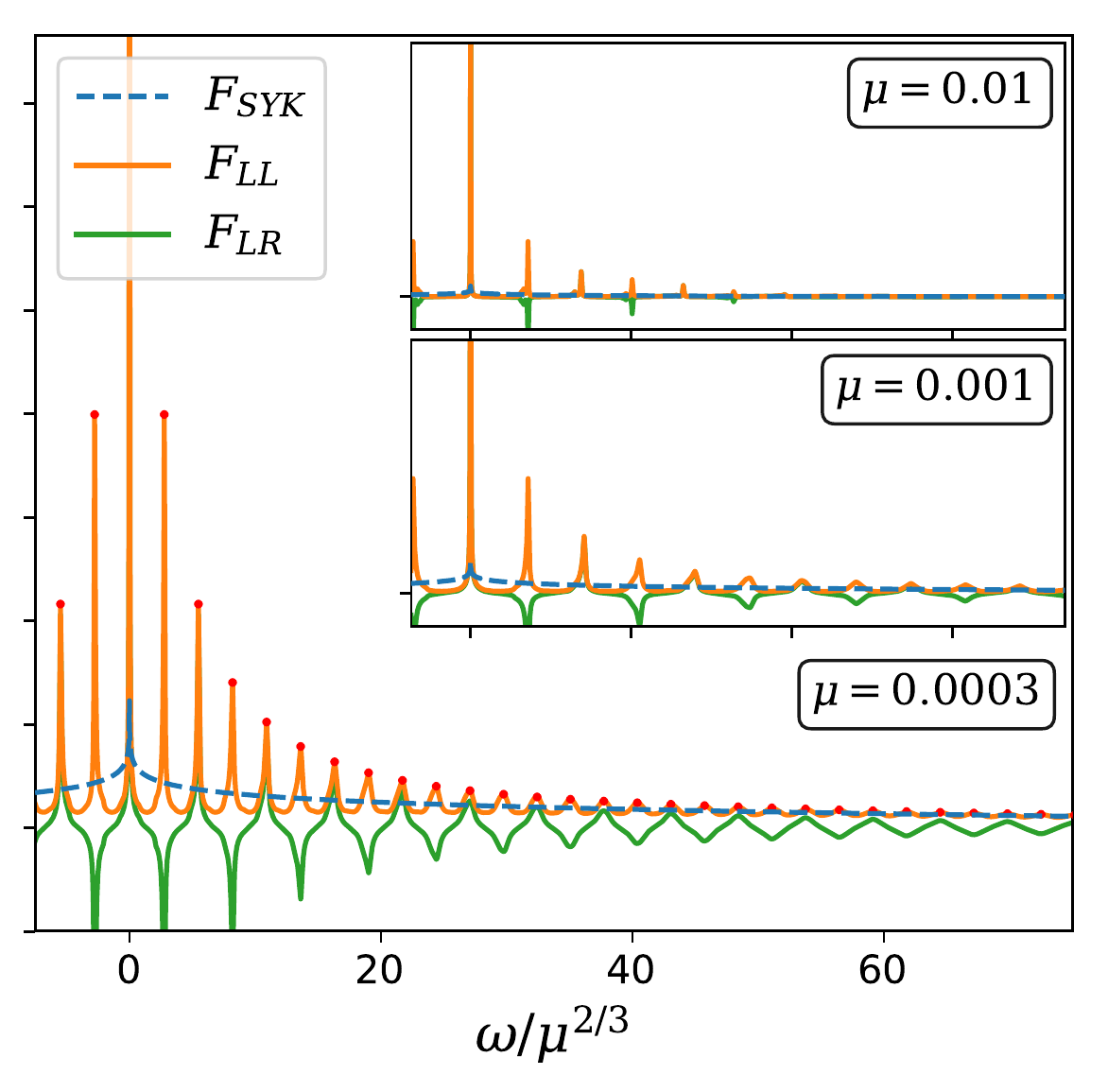}
	\caption{
		Fourier transforms of the transmissions, obtained from the auto-correlation of spectral functions, cf. discussion in text and Eq.~\eqref{eq:Fab}. $F_{LL}$ and $F_{LR}$ show a distinct peak structure that is altogether absent from $F_\mathrm{SYK}$.
		In the revival oscillations of Fig.~3 in main text, slight shifts of the peak frequencies observed here translate to a slow ``beating'' of oscillations.
		\label{fig:SDrevival_FT_SM}}
\end{figure}

Revival dynamics are related to peaks in the spectral function in a subtle way. This is seen by considering transmissions, i.e.\ the square of transmission amplitudes $|T_{\alpha \beta}|^2 = 4G^>_{\alpha \beta}(t) [G^{>}_{\alpha \beta}(t)]^\ast$.
After performing Fourier transforms and representing greater GFs via spectral functions according to
\begin{align}\label{eq:Ggr}
	i G_{\alpha\beta}^>(\omega) = [1-n_F(\omega)]\rho_{\alpha\beta}(\omega),
\end{align}
these can be expressed (at zero temperature) as 
\begin{align}\label{eq:Fab}
	F_{\alpha \beta}( \omega) &= \int d\omega e^{i \omega t} |T_{\alpha \beta}(t)|^2 \\
	&\quad = 8\pi \int_0^\infty d \omega' \rho_{\alpha \beta} (\omega') \rho_{\alpha \beta}(\omega' + |\omega|)~.
	\notag
\end{align}
Thus the positive-frequency auto-correlation of the spectral function gives the Fourier transform of revival dynamics.
The equal energy spacings of spectral peaks (i.e., the conformal tower) obtained in the large-$N$ solution for the spectral function $\rho_{\alpha \beta}(\omega)$ (Fig.~1 of the main text) obviously give rise to sharp features in auto-correlations at multiples of the energy spacing.
As seen in Fig.~\ref{fig:SDrevival_FT_SM}, aside from a dominant peak at zero frequency, the most prominent feature is found at about twice the spectral gap $E_\mathrm{gap}$ in Fig.~1. Further we observe a series of spectral peaks in $F_{\alpha\beta}$, with an even-odd structure inherited from the spectral function, most clearly apparent in the plot of $F_{LR}$ in Fig.~\ref{fig:SDrevival_FT_SM}.
From here, we can associate the revival oscillations in transmission amplitudes to the (almost) equally-spaced peak series in $F_{\alpha\beta}$, with slight deviations from perfectly even spacings leading to the beating observed in Fig.~2 of the main text and Figs.~\ref{fig:SDrevival_SM} in the SM.

\subsection{Thermalization and decay of revivals}

\begin{figure*}[t]
	\includegraphics[width=\textwidth]{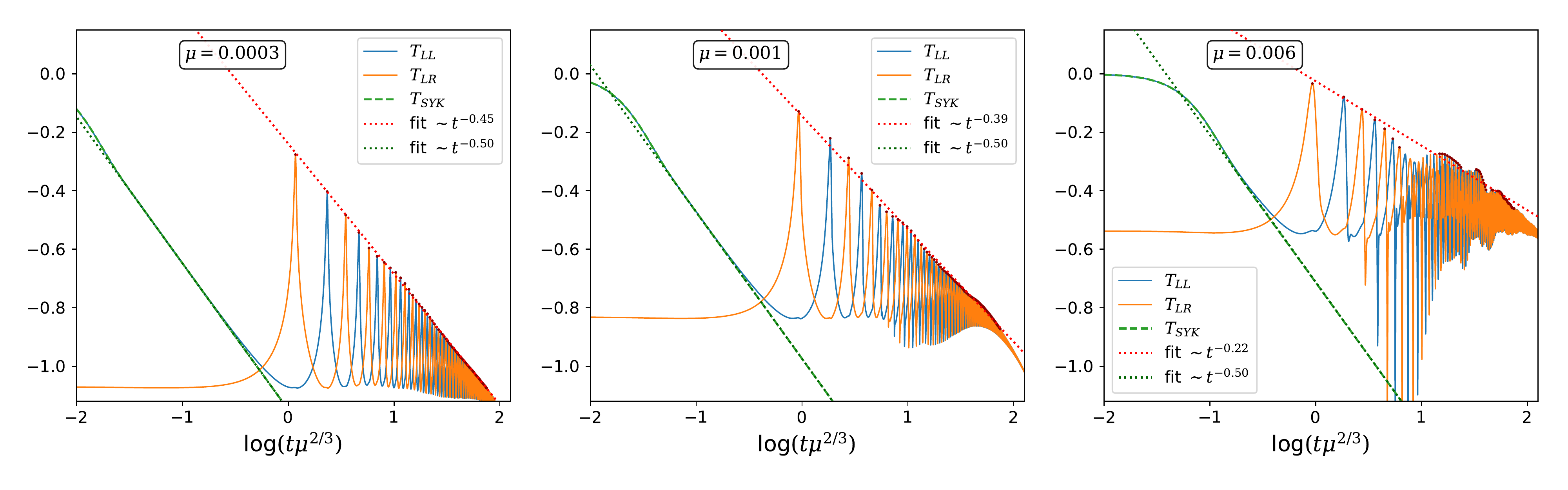}
	\caption{
		Detailed transmission amplitude analysis of revival oscillations at large $N$, cf. Fig.~3 of the main text. We here focus on small $\mu = 0.0003,~0.001,~0.006$ as indicated in the figure panels, where ``beating'' effects are reduced and relatively clean revival oscillations can be observed. In each panel, we indicate and fit the decay of the SYK transmission amplitude $T_\mathrm{SYK}$ (unchanged, modulo shift on $x$-axis), as well as the revival peak heights. The revival peaks used for fitting of the revival amplitude decay are indicated by red dots, and fits are shown as dotted lines.
		\label{fig:SDrevival_SM}}
\end{figure*}

Using the link between the spectral auto-correlation function and the revival dynamics, we now discuss two features of the revivals arising in the large-$N$ solution: ($i$) thermalization of the system -- that is, whether the auto-correlation function $T_{LL}(t) = |G^>_{LL}(t)| \rightarrow 0$ in the infinite-time limit (does the system retain memory of its initial state?), and ($ii$) the decay of revival oscillations (do particles bounce back and forth indefinitely?).

The first question is related to the zero-frequency peak in $F_{LL}(\omega)$. We find that for temperatures $T \ll \mu \ll J$ it has a Lorentzian-like shape with a width $\Gamma$. Thus its Fourier transform contributes $\sim e^{-\Gamma t}$ to the transmission amplitude, with decay rate $\Gamma$.
The system hence eventually thermalizes, on a timescale $\Gamma^{-1}$ that is much longer than both $J^{-1}$ (the timescale for the initial decay of two-point functions) and $\mu^{-2/3}$ (the timescale of revival oscillations). In the zero-temperature limit we expect $\Gamma \rightarrow 0$ when $T\to 0$~\cite{QiZhang2020}, thus the system \emph{fails} to thermalize: a finite, constant value corresponding to the weight in the $\omega=0$ peak remains up to infinite time.

The second question is more subtle, and requires understanding the structure of finite-frequency peaks in $F(\omega)$. For $T \ll \mu \ll J$ those peaks are roughly equally spaced, $\omega_n' = p n$, and resemble Lorentzians with a linearly-increasing width $\Gamma_n = \Gamma_0 n$. For the spectral auto-correlation function at low frequencies, we thus make the ansatz
\begin{equation}
	F_{LL}(\omega) \approx \sum_n \frac{\Gamma_0 n}{(\omega - p n)^2 + (\Gamma_0 n)^2}
\end{equation}
Fourier transforming to real time, we obtain
\begin{align}
	T^2_{LL}(t) &= \int \frac{d\omega}{2 \pi} F_{LL}(\omega) \approx 2 \sum_{n>0} \cos \left(2 \pi p n t \right) e^{- 2 \pi \Gamma_0 n t}
\end{align}
Whenever $t = t_{\rm re} =m/p$ with $m$ integer, all oscillating terms interfere constructively, leading to a sharp revival signal. The height of this revival peak is then given by
\begin{align}
	T^2_{LL}(t_{\rm re}) &\approx 2 \sum_{n>0}  e^{- 2 \pi \Gamma_0 n t}={2\over e^{2 \pi \Gamma_0 t}-1}
\end{align}
When the argument of the exponential is small (valid for $T \ll \mu \ll J$ and time $t$ not too large) one can expand to leading order and find 
\begin{align}
	T^2_{LL}(t_{\rm re}) \simeq \frac{1}{\pi \Gamma_0 t}.
\end{align}
This gives a decay $T_{LL}(t_{\rm re}) \sim t^{-1/2}$ similar to the observation in Fig.~2 and~\ref{fig:SDrevival_SM}. Interestingly this decay has the same power as the uncoupled SYK model, leading us to hypothesize that the SYK interactions are responsible for the $\Gamma_n \sim n$ broadening of the finite-frequency peaks.

This argument is supported by the revival oscillations in Fig.~\ref{fig:SDrevival_SM} (and Fig.~2 of the main text). Our first observation is that we reliably recover the expected two-point function decay of SYK, where $G_\mathrm{SYK}(t)\sim t^{-1/2}$.
This behavior is reflected in the initial trend of the transmission amplitude $T_{LL}$ up to times $\sim \mu^{-2/3}$ where the wormhole physics in the form of revivals arise. For small enough $\mu$ and in the absence of ``beating'' in the revival oscillation, we can fit the decay of transmission amplitudes to a power law,  $T_{LL,LR}(t = t_\mathrm{re}) \sim t^{-\nu}$. The exponent $\nu$ is found to  approach the characteristic value for the decay of correlation functions in SYK, $\nu \simeq \frac12$.

This decay of revivals goes beyond the predictions that the dual gravity theory has made so far, and possibly necessitates an additional mechanism beyond the low-energy conformal field theory of Maldacena and Qi~\cite{Qi2018}. 
A detailed analysis of the decay of revivals may provide valuable hints towards an extension of the low-energy theory -- e.g. to allow excitations to interact with a finite density of ``bulk particles'' and to dissipate energy as they traverse the wormhole.
\footnote{We thank J. Maldacena and A. Milekhin for clarifying to us that this dissipative mechanism can be included in the gravitational description of the MQ model.}

\subsection{Symmetries of the Maldacena-Qi model}
\label{app:sym}

In this section we discuss useful symmetries of the MQ model. The most obvious is conservation of fermion parity, $P = (-i)^N \prod_{j=1}^N (\chi^j_L \chi^j_R)$, which is equivalent to fermion number (or charge) $Q~\text{mod}~2$. This clearly commutes with the Hamiltonian~(1) because the latter has only two-fermion and four-fermion terms. The MQ model also has a $Z_4$ symmetry defined as $q \equiv Q~\text{mod}~4$, which crucially relies on the perfectly correlated disorder between $L$ and $R$ SYK Hamiltonians in Eqs.(1)-(2) \cite{Garcia2019}.

The coupling term in the MQ Hamiltonian is proportional to the charge $Q = \sum_j c_j^\dagger c_j$. However it does not commute with the SYK terms (only $Q~\text{mod}~4$ does). This is the main difference between the MQ and complex SYK models, the latter having $U(1)$ charge symmetry~\cite{Sachdev2015}.

We can also define a unitary operator $U$ which takes $\chi_L^j \rightarrow \chi_L^j$ and $\chi_R^j \rightarrow -\chi_R^j$. This is a symmetry of the model with zero coupling ($\mu=0$), and it anti-commutes with the charge operator, $\{Q, U\} = 0$. As a result the $Z_4$ sectors $q=1,3$, which are related by $Q \leftrightarrow -Q$, are mapped to each other by $U$ and thus degenerate, $\ket{n,3} = U \ket{n,1}$ and vice versa. This will be useful in the following.

\subsection{Analytical result for revivals at small coupling}

To better understand the revival dynamics, it is helpful to look at the structure of the ``single-Majorana'' excitations that we wish to transfer from one side to the other:
\begin{equation}
	\ket{j,L} \equiv \chi_j^L \ket{0} \quad , \quad \ket{j,R} \equiv \chi_j^R \ket{0}
\end{equation} 
Expanding in the eigenbasis of the Hamiltonian, we have
\begin{equation}
	\ket{j,\alpha} = \sum_{n,q} c_j^\alpha(n,q) \ket{n, q}
\end{equation}
with $\alpha=L,R$. Here the sum is taken over all eigenstates $\ket{n}$ in the $Z_4$ sectors $q=0,1,2,3$. The ground state $\ket{0}$ of the MQ model is always in the $q=0$ sector, therefore the excited states defined above are in one of the two odd-parity sectors, $q=1, 3$. The coefficients $c_j^\alpha(n,q)$ thus vanish for $q=0, 2$.

The (unaveraged) causal Green's functions $G^>_{LL}(t) = \langle 0 | \chi_j^L(t) \chi^L_j | 0 \rangle $ and $G^>_{LR}(t) = \langle 0 | \chi_j^L(t) \chi^R_j | 0 \rangle$ thus read
\begin{align}
	G^>_{LL}(t) &= e^{iE_0 t} \sum_n \sum_{q=1,3} |c_j^L(n,q)|^2 e^{-iE_{n,q}t} \\
	G^>_{LR}(t) &= e^{iE_0 t} \sum_n \sum_{q=1,3} \left( c_j^L(n,q) \right)^* c_j^R(n,q) e^{-iE_{n,q} t} \nonumber
\end{align}
for a given, fixed mode $j$. Due to the all-to-all, random nature of the interactions within each flake, any choice of $\chi_j$ is equivalent on average -- we thus choose to work with a single mode $j$ in exact diagonalization for simplicity. To understand the transmission of those ``single-particle" excitations from one side of the wormhole to the other, we consider the transmissions
\begin{align}
	T^2_{LL}(t) &= \sum_{m,n} \sum_{p,q} |c_j^L(m,p)|^2 |c_j^L(n,q)|^2 e^{i(E_{m,p}-E_{n,q})t}, \\
	T^2_{LR}(t) &= \sum_{m,n} \sum_{p,q} \left[c_j^R(m,p) c_j^L(n,q) \right]^* c_j^L(m,p) c_j^R(n,q)  \nonumber \\
	&\times e^{i(E_{m,p}-E_{n,q})t}.
\end{align}
For the smallest system size $2N=16$, and in the limit $\mu \ll \lambda \ll 1$, where $\lambda$ is the typical finite-$N$ level spacing, two dramatic simplifying features occur: 

1) The spectrum in the odd sectors ($q=1, 3$) acquires the simple form
\begin{equation}
	E_{m,1} = E_{m,3} + \omega_D.
	\label{eq:SM_uniformshift}
\end{equation}
There is a \emph{uniform} spectral shift between the two charge sectors. This is the mechanism at the root of the revival dynamics (and the splitting of doublets in the spectral function) observed in ED. A simple calculation given below reveals why that uniform shift occurs.

\begin{figure*}[t]
	\includegraphics[width =\textwidth]{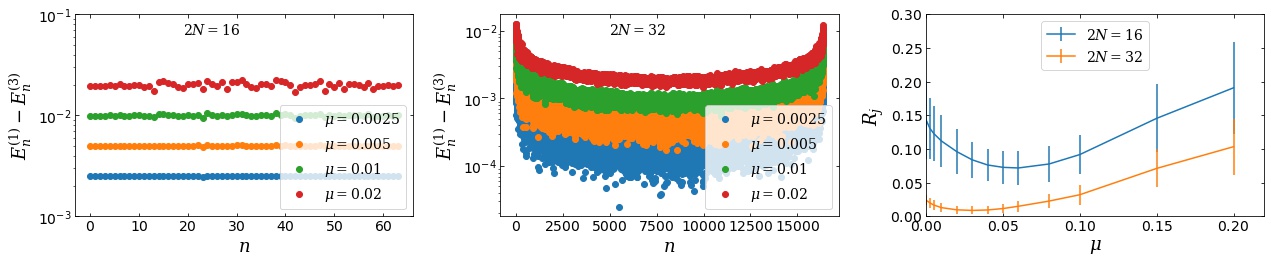}	
	\caption{Additional ED results. (a) and (b): Difference between the spectra in the odd $Z_4$ symmetry sectors, $E_{n,1} - E_{n,3}$, as a function of $\mu$ for (a) $2N=16$ and (b) $2N=32$. We observe a uniform spectral shift for $\mu \ll \lambda$ in (a) but not in (b). (c) Participation ratio $R_j$ of the excitation $\chi^j_R \ket{0}$ for various $\mu$.
		\label{fig:suppl_ED}}
\end{figure*}

2) The absolute value of the coefficients $c_j^{\alpha}(n,q)$ becomes \emph{independent} of the charge sector $q$, and respect the following relation for all $n$:
\begin{equation}
	c_j^L(n,1) = i c_j^R(n,1) \; , \; c_j^L(n,3) = -i c_j^R(n,3) 
\end{equation}
Combining these two features, we obtain the simple result
\begin{align}
	|G^>_{LL}(t)|^2 &= 2 \left( 1 + \cos \omega_D t \right) G_0(t) \nonumber \\
	|G^>_{LR}(t)|^2 &= 2 \left( 1 - \cos \omega_D t \right) G_0(t)
	\label{eq:transmissions_SM}
\end{align}
where
\begin{equation}
	G_0(t) = \sum_{m,n} |c_j(m)|^2 |c_j(n)|^2 e^{i(E_{m,1}-E_{n,1})t}
	\label{eq:G0}
\end{equation}
In this limit the behavior of the revival dynamics becomes clear: there is an envelope $\sim (1 \pm \cos \omega_D t)$ at a slow frequency set by the splitting between doublets, and fast oscillations at a large number of non-universal frequencies $E_{m,1} - E_{n,1}$ contained in $G_0(t)$, set by finite-size energy gaps characteristic of SYK physics. Considering only the non-oscillating part with $m=n$, we get 
\begin{equation}
	G_0(t) = \sum_m |c_j(m)|^4 \equiv R_j
\end{equation}
where we defined the participation ratio $R_j$ of the initial excitation. This quantity characterizes the ``inverse number of eigenstates" that are  involved in the dynamics and sets the amplitude of the revivals,
\begin{equation}
	|G^>_{LL/LR}(t)|^2 = 2 \left( 1 \pm \cos \omega_D t \right) R_j.
\end{equation}
It rapidly decreases with system size $N$, as shown in Fig.~\ref{fig:suppl_ED} c. This highlights the fact that the mechanism presented here cannot account for the revivals obtained in the large-$N$ solution -- it relies crucially on the hierarchy of energy scales $\mu \ll \lambda \ll 1$. 

Another way to understand this is to realize that Eq.~(\ref{eq:G0}) corresponds to the Green's function of the canonical SYK model. As $N$ increases, this quantity should approach the power-law decay $G_0(t) = |G_{\rm SYK}(t)|^2 \sim |t|^{-1}$. This decay occurs on a timescale $J^{-1}$ much \emph{smaller} than the revivals and will thus completely obscure them. What saves the day for small $N$ is that the decay of $G_{\rm SYK}(t)$ is cut-off by finite-size effects at time $t \sim \lambda^{-1} \sim e^N/N$, which leaves room for revival oscillations at longer times, as seen in Fig.~4 of the main text.

For larger $N$ (such as $2N=32$) the uniform splitting result, Eq.~(\ref{eq:SM_uniformshift}) does not hold exactly -- there is a weak dependence of the splitting frequency on the eigenstate index $n$, as shown in Fig.~\ref{fig:suppl_ED}. This gives rise to a decay of revivals not captured by Eq.~\eqref{eq:transmissions_SM}, as observed in Fig.~4 of the main text, because the doublets oscillate with a distribution of frequencies. Furthermore, this analysis only applies to system sizes $N~\text{mod}~8= 0$ which correspond to the Gaussian Orthogonal Ensemble (GOE) of random matrix theory, where energy levels of the SYK model are non-degenerate. For other system sizes (with GUE or GSE statistics) the SYK spectrum has additional degeneracies which renders our solution invalid -- for example, doublet splittings as in Eq.~\eqref{eq:SM_uniformshift} are generally not observed for these system sizes.  

\subsection{Perturbation theory}

In the very weak coupling regime $\mu \ll \lambda$, one can utilize degenerate perturbation theory to gain some insight. As argued above, for $\mu=0$ the spectrum of the model in identical in the charge sectors $q=1,3$ due to the correlated disorder between the two SYK models. Introducing a non-zero $\mu$ couples each pair of degenerate states. To first-order we get a shift (for $U,~Q$ see also discussion  of symmetries of the MQ model above)
\begin{align}
	E_{n,1} &= \mu \bra{n,1} Q \ket{n,1} \\
	E_{n,3} &= \mu \bra{n,3} Q \ket{n,3} = \mu \bra{n,1} U^\dagger Q U \ket{n,1} \nonumber \\
	&= - \mu \bra{n,1} Q \ket{n,1} = -E_{n,1}
\end{align}
which is \emph{opposite} in the two $Z_4$ sectors. An additional result, valid for $2N=16$ only, is that the expectation value 
\begin{align}\label{eq:N16magic}
	E_{n,1} &= \mu \bra{n,1} Q \ket{n,1} = \frac{\mu}{2}
\end{align}
is independent of the eigenstate index $n$ (see Fig.~\ref{fig:suppl_ED}). This fixes $E_{m,1} = E_{m,3} + \omega_D$ above with $\omega_D = \mu$. However, we have not yet been able to find a solid argument to explain the statement of Eq.~\eqref{eq:N16magic}.

\end{document}